\documentclass{aa}  

\usepackage{graphicx}

\usepackage{xcolor}

\usepackage{txfonts}
\usepackage[colorlinks, allcolors=blue]{hyperref}
\def\Lya     {\ensuremath{\text{Ly}\alpha}}

\def\oiiibr    {\ensuremath{\text{{[O\,\textsc{iii}]}}}}

\def \rcrosscorr {$r_{0}^{\textrm{QG}}\approx7.60_{-1.61}^{+1.65}\,\textrm{h}^{-1}\,\textrm{cMpc}$}

\def \rautocorr {$r_0^{\textrm{QQ}}\approx12.61_{-4.53}^{+5.72}\,\textrm{h}^{-1}\,\textrm{cMpc}$}

\def \logMmin {$\log_{10}(M_{\textrm{halo, min}}/\textrm{M}_{\odot})= 11.59_{-0.70}^{+0.59}$}

\def \lognhalo {$\log_{10}(n_{\textrm{halo, min}}/\textrm{cGpc}^{-3})=2.83_{-2.40}^{+1.95}$}

\def\fD {$\log_{10}(f_{\textrm{duty}})\approx-3.36_{-1.95}^{+2.40}$}

\def\tQ {$\log_{10}(t_{\textrm{QSO}}/\textrm{yr})\approx5.48_{-1.95}^{+2.40}$}

\begin{document}

   \title{A first look at quasar-galaxy clustering at $z\simeq7.3$}

   \author{Jan-Torge Schindler\inst{1} 
           \and Joseph F. Hennawi\inst{2}\inst{3}
           \and Frederick B. Davies \inst{4}
           \and Sarah E. I. Bosman \inst{5,4}
           \and Feige Wang \inst{6} 
           \and Jinyi Yang \inst{6}          
           \and Anna-Christina Eilers \inst{7,8}
           \and Xiaohui Fan \inst{9}
           \and Koki Kakiichi \inst{10}
           \and Elia Pizzati \inst{2}
           \and Riccardo Nanni \inst{2}
          }

\institute{Hamburger Sternwarte, Universität Hamburg, Gojenbergsweg 112, D-21029 Hamburg, Germany\\
\email{jan-torge.schindler@uni-hamburg.de}
\and Leiden Observatory, Leiden University, P.O. Box 9513, 2300 RA Leiden, The Netherlands
\and Department of Physics, Broida Hall, University of California, Santa Barbara, Santa Barbara, CA 93106-9530, USA
\and Max-Planck-Institut f\"{u}r Astronomie, K\"{o}nigstuhl 17, 69117 Heidelberg, Germany
\and Institute for Theoretical Physics, Heidelberg University, Philosophenweg 12, D–69120, Heidelberg, Germany
\and Department of Astronomy, University of Michigan, 1085 S. University Ave., Ann Arbor, MI 48109, USA
\and Department of Physics, Massachusetts Institute of Technology, Cambridge, MA 02139, USA
\and MIT Kavli Institute for Astrophysics and Space Research, Massachusetts Institute of Technology, Cambridge, MA 02139, USA
\and Steward Observatory, University of Arizona, 933 N Cherry Ave, Tucson, AZ 85721, USA
\and Cosmic Dawn Center (DAWN), Niels Bohr Institute, University of Copenhagen, Jagtvej 128, DK-2200 København N, Denmark
}


\abstract
{
Linking quasars to their dark matter environments provides critical insights into the formation and early growth of supermassive black holes (SMBHs).
We present JWST observations of the environments surrounding two high-redshift quasars,  J0252$-$0503 at $z = 7.0$ and J1007$+$2115 at $z = 7.5$,  which enable the first constraints on quasar–galaxy clustering at $z \simeq 7.3$.
Galaxies in the vicinity of the quasars were selected through ground-based and JWST/NIRCam imaging and were then spectroscopically confirmed with JWST/NIRSpec using the multi-shutter assembly (MSA). 
Over both fields, we identified 51 $z\!>\!5$ galaxies, of which eight are found within a $\Delta v_{\textrm{LOS}}=\pm1500\,\rm{km}\,\rm{s}^{-1}$ line-of-sight velocity window from the quasars and another eight in the background.
The galaxy J0252\_8713, located just $7\,\rm{pkpc}$ and $\Delta v_{\textrm{LOS}} \approx 360\,\rm{km}\,\rm{s}^{-1}$ from quasar J0252$-$0503, emerges as a compelling candidate for one of the most distant quasar-galaxy mergers.
Combining the galaxy discoveries over the two fields, we measured the quasar-galaxy cross-correlation and obtain a correlation length of $r_0^{\rm{QG}}\approx7.6_{-1.6}^{+1.7}\,h^{-1}\,\rm{cMpc}$, based on a power-law model with a fixed slope of $\gamma_{\rm{QG}} = 2.0$.
Under the assumption that quasars and galaxies trace the same underlying dark matter density fluctuations, we infer a minimum dark matter halo mass for $z\simeq7.3$ quasars of $\log_{10}(M_{\textrm{halo, min}}/\textrm{M}_{\odot})= 11.6_{-0.7}^{+0.6}$ in a halo model framework.
Compared to measurements from EIGER at $\langle z \rangle = 6.25$ and ASPIRE at $\langle z \rangle = 6.7$ (where $\log_{10}(M_{\textrm{halo, min}}/\textrm{M}_{\odot}) \gtrsim 12.1$), our clustering results provide tentative evidence for a nonmonotonic redshift evolution of quasar clustering properties. 
We further estimate a quasar duty cycle of $f_{\rm{duty}}\approx0.05\%$, consistent with constraints from quasar proximity zones and intergalactic medium (IGM) damping wings. 
However, this implies very short phases of quasar activity, exacerbating the challenge to build billion solar mass SMBHs in only $700\,\rm{Myr}$ of cosmic time. 
}

   \keywords{galaxies: high-redshift -- quasars: general -- quasars: supermassive black holes --  large-scale structure of Universe
               }

   \maketitle
%

\section{Introduction}
Observations of high-redshift quasars probe the early UV-luminous growth phases of supermassive black holes (SMBHs). 
They have shown that billion solar mass SMBHs are already in place at redshifts of $z\gtrsim6$,  hence leaving less than one billion years for their assembly \citep[see][for recent reviews]{Fan2023, Inayoshi2020}.
Particularly, quasars at the redshift frontier ($z \simeq 7.5$), observed just 700\,Myr after the Big Bang \citep{Banados2018, YangJinyi2020, WangFeige2021}, pose a significant challenge to the standard model of supermassive black hole growth via Eddington-limited accretion from stellar-mass seeds, assuming a radiative efficiency of $\epsilon = 0.1$.
This tension and constraints on the quasar duty cycles \citep[e.g.,][]{Eilers2017, Davies2019, Eilers2024} motivate episodes of more rapid growth than in the standard model
\citep[e.g.,][]{Volonteri2005, Ohsuga2005, Madau2014, JiangYanFei2019}, possibly in combination with very massive initial seeds \citep[$M_{\rm{seed}}\gtrsim10^4\,\rm{M}_\odot$;][]{Omukai2001, Bromm2003, Devecchi2009}.
The immediate environments, which either provide exotic conditions for the genesis of massive seeds or act as reservoirs for rapid gas accretion, play a critical role in SMBH formation and early growth. 
Hence, building a comprehensive understanding of SMBHs in a cosmological context requires linking the instances of SMBH growth to their environments and ultimately to large-scale structure evolution.

Cosmological simulations \citep[e.g.,][]{DiMatteo2005, Khandai2015, Feng2016} can produce $10^9\,\rm{M}_\odot$ SMBHs similar to the observed $z\sim6$ quasar population starting from massive seeds ($M_{\rm{seed}}\gtrsim10^4\,\rm{M}_\odot$).
These systems are hosted in massive star-forming galaxies ($M_{\star}\gtrsim10^{11}\,\rm{M}_\odot$) residing in rare $M\gtrsim10^{12}\,\rm{M}_\odot$ dark matter halos, and thus probe the most overdense regions of the Universe \citep[e.g.,][]{DiMatteo2012, Costa2014, Khandai2015, Feng2016, Barai2018, Valentini2021}. 
As a consequence, the $\Lambda\rm{CDM}$ cosmological model predicts a high number of companion galaxies around these SMBHs \citep[e.g.,][]{Munoz2008, Tinker2010}.

For many years, both ground-based and space-borne imaging campaigns led to inconclusive or even contradictory results regarding the environments of high-redshift quasars.
Using multi-band photometry, galaxy candidates were either selected using the Lyman-break technique or as Lyman-$\alpha$ emitters (LAEs) in narrowband filters.
These techniques led to numerous claims of overdensities around $z\gtrsim6$ quasars \citep[e.g.,][]{Stiavelli2005, Zheng2006, Kim2009, Ajiki2006, Utsumi2010, Morselli2014, Pudoka2024} with as many studies reporting average environments \citep[e.g.,][]{Willott2005, Banados2013, Simpson2014, Goto2017, Mazzucchelli2017}.
Recent studies utilized VLT/MUSE to identify LAEs spectroscopically in the quasar environment \citep[][]{Mignoli2020, Meyer2022} or ALMA to detect galaxies via 
{\ensuremath{\text{[C\,\textsc{ii]}}}}$_{\rm{158\,\mu\rm{m}}}$ emission \citep{Champagne2018, Miller2020, Meyer2022}. 
With small sample sizes and diverse galaxy identification techniques, the mixed results precluded the emergence of a coherent picture. 

Over the last two years the spectroscopic capabilities of the James Webb Space Telescope \citep[JWST;][]{Gardner2023} have proven invaluable in the study of quasar environments. 
The Emission-line Galaxies and Intergalactic Gas in the Epoch of Reionization \citep[EIGER;][]{Kashino2023} and A SPectroscopic Survey of Biased Halos in the Reionization Era \citep[ASPIRE;][]{WangFeige2023} were the first survey projects to systematically use NIRCam Wide Field Slitless Spectroscopy (WFSS) to identify  \oiiibr-emitting galaxies around $z\gtrsim6$ quasars.
\citet{Kashino2023} and \citet{WangFeige2023} were also the first to find high cMpc-scale overdensities of \oiiibr-emitters around quasars J0100+2802 ($z=6.3$) and J0305-3150 ($z=6.6$), respectively. 
The quasar two-point correlation function has proven to be a powerful tool for connecting the clustering properties of quasars with their hosting dark matter halos \citep[e.g.,][]{Shanks1994, Croom2001}. 
At low to intermediate redshifts, $1 \le z \le4$, auto-correlation function analyses on large quasar samples find them to be hosted in massive dark matter halos ($\sim10^{12}\,\rm{M}_\odot$) \citep[e.g.,][]{Porciani2004, Croom2005, Porciani2006, ShenYue2007, Ross2009, White2012, Eftekharzadeh2015, Laurent2017, HeWanqiu2018}.
At $z\sim1$ this mass is close to the characteristic mass of the halo mass function. However, with increasing redshift $\sim10^{12}\,\rm{M}_\odot$ dark matter halos become increasingly rare and are highly biased regions of the Universe.

Due to the rapid decline in the quasar volume density at high redshifts \citep[e.g.,][]{Matsuoka2018, Wang2019, Schindler2023, Matsuoka2023}, it is an extraordinary challenge to build large enough quasar samples at $z\gtrsim6$ for an auto-correlation analysis. 
Exploiting the large number of faint quasar discoveries of the Subaru High-z Exploration of Low-Luminosity Quasars (SHELLQs) program, \citet{Arita2023} compiled a sample of 107 quasars over an area of $891\,\rm{deg}^2$ for a first quasar auto-correlation analysis at $z\sim6$.
The authors report a high dark matter halo mass of $5_{-4.0}^{+7.4}\times10^{12}\,\rm{M}_\odot$ and a bias parameter of $b=20\pm8.7$ \citep[but see Appendix C in][]{Pizzati2024}. Although there are considerable uncertainties, their results reflect a continuation of the lower-redshift trend. The dark matter halo mass remains high, while the bias parameter increases with redshift. 
This leads the authors to suggest that quasars activate at a characteristic dark matter halo mass of $\sim10^{12}\,\rm{M}_\odot$ across redshifts.

An alternative pathway to constrain the clustering properties of quasars is to utilize the quasar-galaxy cross-correlation function. Under the assumption that quasars and galaxies trace the same underlying dark matter density distribution, their cross-correlation function is determined by their respective auto-correlation functions. 
Hence, by determining the quasar-galaxy cross-correlation and the galaxy auto-correlation functions, one can infer the quasar auto-correlation, circumventing the limitations of small quasar sample sizes. 
Exploiting the spectroscopic capabilities of JWST, \citet{Eilers2024} used the EIGER survey \citep{Kashino2023, Matthee2023} to systematically study the clustering of $\langle z\rangle=6.25$ \oiiibr-emitting galaxies in four bright quasar fields. The authors simultaneously constrain the galaxy auto-correlation and the quasar-galaxy cross-correlation functions to conclude that quasars at $z\approx6$ on average reside in massive $\approx10^{12.4}\,\rm{M}_\odot$ halos. 
They also note that the number of companion galaxies varies significantly from field to field.
\citet{Pizzati2024} applied a novel inference framework \citep{Pizzati2024a} that combines the empirical results on the correlation functions of \citet{Eilers2024} and constraints on the quasar luminosity function with state-of-the-art dark-matter-only cosmological simulations. At $z\sim6$ the authors confirm massive quasar host dark matter masses of $\log_{10}(M_{\textrm{halo, min}}/\textrm{M}_{\odot}) \approx 12.5$ in contrast to the \oiiibr-emitting galaxies galaxy 
population ($M_{\rm{DM}}\approx10^{10.9}\,\rm{M}_\odot$). 
\citet{HuangJiamu2026} \citep[also see][]{WangFeige2026} measure the quasar-galaxy cross-correlation at $z\approx6.7$ based on \oiiibr-emitting galaxies in 25 quasar fields from the ASPIRE survey \citep{WangFeige2023}. The authors infer an average dark matter halo mass of $\log_{10}(M_{\textrm{halo, min}}/\textrm{M}_{\odot}) \approx 12.13$. 
Additionally, the field-to-field variation in the number of companion galaxies, first observed by \citet{Eilers2024}, is confirmed with the larger ASPIRE sample \citep{WangFeige2026}. This suggests that the previously inconclusive overdensity measurements around quasars may be in part the result of cosmic variance. 

These recent studies have shown the feasibility of JWST observations to extend quasar-galaxy clustering measurements within the first billion years of cosmic time, a critical time for the early growth of SMBHs. 
However, even with significant efforts, the combination of the decreasing quasar volume density \citep{Wang2019, Matsuoka2023} and the limitations of existing wide-area surveys have resulted in less than a dozen $z\gtrsim7$ quasar discoveries \citep[e.g.,][]{Mortlock2011, WangFeige2018, Matsuoka2019, Matsuoka2019a, YangJinyi2019, YangJinyi2020, WangFeige2021, Banados2025, Matsuoka2025a}. 
Recent investigations for companion galaxies around some of the highest-redshift quasars have shown a diversity of environments \citep{RojasRuiz2024, Pudoka2025}.

In this work we present an environmental study of two $z\gtrsim7$ quasar fields leading to first constraints on the $z\simeq7.3$ quasar-galaxy cross-correlation function. Our work is based on the JWST observations of the Cycle 1 GO program 2073 (PI: Hennawi) “Towards Tomographic Mapping of Reionization Epoch Quasar Light-Echoes”.
In Section\,\ref{sec:observations} we discuss the observations and the subsequent data reduction. The galaxy discoveries in the two quasar fields are discussed in Section\,\ref{sec:galaxies}, and we present our analysis of quasar-galaxy clustering in Section\,\ref{sec:clustering}. 
We continue to discuss the interpretation of our results in context with the recent literature in Section\,\ref{sec:discussion} and conclude our work in Section\,\ref{sec:summary}.
All magnitudes are provided in the AB system, and we adopt a concordance cosmology \citep[e.g.,][]{Hinshaw2013} with $H_{0}=70\,\textrm{km}\,\textrm{s}^{-1}\,\textrm{Mpc}^{-1}$, $\Omega_{\Lambda}=0.7$, and $\Omega_{M}=0.3$ for cosmological calculations.

\section{Observations}\label{sec:observations}
The clustering analysis presented in this work is based on the JWST program GO 2073 (PI: J. Hennawi). This program was designed to spectroscopically identify galaxies in the surroundings and background of two $z>7$ quasars, J1007$+$2115 at $z=7.51$ \citep[$M_{1450}\!=\!-26.62\,\textrm{mag}$,][]{YangJinyi2020} and J0252$-$0503 at $z=7.00$ \citep[$M_{1450}\!=\!-25.77\,\textrm{mag}$,][]{YangJinyi2019, WangFeige2020}. A future program aims to use the background galaxies to tomographically map the ionized region around those quasars, known as their light echos \citep{Schmidt2019, Eilers2025}, for constraints on their active lifetimes, whereas the galaxies at the quasar redshift were targeted to also provide first constraints on quasar-galaxy clustering at $z\simeq7.3$.

Within JWST Cycle\,1 we collected JWST/NIRCam photometry in the two quasar fields as the basis for our galaxy candidate selection and then spectroscopically followed-up these candidates with the NIRSpec/MSA in the same cycle. 
\subsection{JWST/NIRCam and ground-based imaging}
The JWST/NIRCam observations provide imaging in four filters (F090W, F115W, F277W, F444W) over an area of $\sim 5'\times6'$ centered around the target quasar. Each field consists of two pointings with a \texttt{FULLBOX 6} dither pattern and a \texttt{MEDIUM 8} readout pattern to build two larger mosaics. 
The JWST imaging observations in the J1007$+$2115 field took $14952\,\rm{s}$ science time with $3736.4\,\rm{s}$ per exposure and SW/LW filter pair.
As the observations in the J0252$-$0503 field targeted slightly lower-redshift sources, we used $2512.4\,\rm{s}$ per exposure and SW/LW filter pair for a total science time of $10056\,\rm{s}$ to reach similar signal-to-noise constraints.
We downloaded the data using the \texttt{jwst\_mast\_query}\footnote{\url{https://github.com/spacetelescope/jwst_mast_query}} Python package and reduced it with version 1.6.3 of the JWST Science calibration pipeline\footnote{\url{https://jwst-pipeline.readthedocs.io/en/latest}} (\texttt{CALWEBB}; CRDS context \texttt{jwst\_1046.pmap}).
A range of additional steps were added to the default data reduction, which are described in detail in the methods section of \citet{Schindler2025}.

We extracted source photometry using the SExtractor \citep{Bertin1996} software on the resampled NIRCam mosaics, convolved to the lower resolution of the F444W filter. 
We performed source detection on an inverse variance weighted signal-to-noise image stack of the four filter mosaics. For all detections we calculated Kron \citep{Kron1980} aperture photometry using a Kron parameter of 1.2. Standard aperture flux corrections were made, adopting the procedures outlined in \citet{Bouwens2016}. Additionally, we corrected for extended point-spread-function (PSF) wings using empirically generated PSFs based on point sources in the field. 

In order to improve our high-redshift galaxy selection we obtained deep ground-based dropout images. We supported the J1007$+$2115 field galaxy selection with LBT/LBC \citep{Giallongo2008} r-band and i-band photometry, whereas we added Keck/LRIS \citep{Oke1995, McCarthy1998, Rockosi2010} G- and I-band photometry to the J0252$-$0503 field.
Both LBT/LBC and Keck/LRIS imaging reduction were performed with a custom data reduction pipeline named {\tt PyPhot}.\footnote{\url{https://github.com/PyPhot/PyPhot}} For the LBT/LBC imaging reduction, we refer to the methods section of \citet{Schindler2025}. 
The reduction steps for the LRIS imaging data are similar to those for the LBC imaging data. Briefly, the main calibrations include bias subtraction, flat-fielding, fringing removal (only for the $I$ band), and sky background subtraction. The master bias and flat frames were constructed using a sigma-clipped median combination of a series of bias and sky flats, respectively. The master fringe image in the $I$ band was created by combining all science exposures in that band after masking out objects detected with {\tt SExtractor} \citep{Bertin1996}. Cosmic ray and other outlier rejections were performed using the Laplacian edge detection algorithm \cite{vanDokkum2001}. After the imaging calibration, we performed an initial bright point source detection with {\tt SExtractor}, which was used to align all the individual images to GAIA DR3. We calibrated the photometric zero points using well-detected point sources that have counterparts in the Pan-STARRS photometric catalog \cite{Chambers2016}. Finally, after the astrometric and photometric calibration, the mosaics for each band were produced using {\tt SCAMP} \citep{Bertin2006} and {\tt Swarp} \citep{Bertin2002}.

\subsection{Galaxy candidate selection}\label{sec:gal_cand_sel}
We targeted high-redshift galaxies that are either in the background or at the redshift of the quasar. Our minimum target redshift was designed to include galaxies $3000\,\textrm{km}\,\textrm{s}^{-1}$ in the foreground of the quasar, whereas the maximum target redshift was chosen so that the quasar is still within the Ly$\beta$ forest of the background galaxy. In practice, this results in redshift ranges of $7.43 < z < 9.09$ and $6.92 < z < 8.48$ for the J1007$+$2115 and J0252$-$0503 field, respectively. 

Our galaxy candidate selection began with standard signal-to-noise ratio (S/N) requirements of $2.0$ for source detections in the F115W and F277W bands. Informed by the photometric properties of mock galaxies in the JAGUAR catalog \citep{Williams2018}, we applied a color cut, $\|\rm{m_{\rm{F115W}}}-\rm{m_{\rm{F277W}}}\| > 1.5$, to reduce contaminants.
Galaxies at the target redshift were selected via the Lyman-break using a probabilistic dropout selection on the F090W to F115W color. For each color value we assigned a purity value based on the JAGUAR mock galaxies. Here, purity refers to the fraction of mock galaxies with a given color within the target redshift range compared to all mock galaxies with the same color value. After each galaxy candidate had been assigned a purity value, we ranked all candidates by descending purity. 
In order to design the NIRSpec/MSA follow up observations, we grouped candidates in six different priority classes based on their purity rank. The first 100 sources were assigned to class 1. Classes 2, 3, 4, and 5 are then filled with the next 100, 100, 700, and 1000 candidates in our list. All remaining candidates were assigned to class 6. These priority classes constituted the input for the MSA mask design using the eMPT tool \citep{Bonaventura2023}.

Before the NIRSpec/MSA masks were generated, we performed rigorous visual inspection of galaxy candidates. For that purpose, we calculated photometric redshifts with \texttt{bagpipes}\footnote{\url{https://bagpipes.readthedocs.io/en/latest/}} \citep{Carnall2018} based on the ground-based and JWST photometry. 
We visually inspected all sources in the first five priority classes. Sources where the photometry had been seriously impacted by image artifacts were removed. In cases where a significant F090W flux was detected that was incompatible with a $z\gtrsim6.9$ galaxy (based on the photometric redshifts) we demoted the candidate to the next lower priority class. This procedure prioritizes sources in the targeted redshift range without a strict low-redshift cutoff. 
The galaxy selection procedure was carried out independently for the two quasar fields, which have two different target redshift ranges. 

\subsection{Spectroscopic follow-up observations}
\begin{figure*}[ht!]
    \centering
    \includegraphics[width=0.8\linewidth]{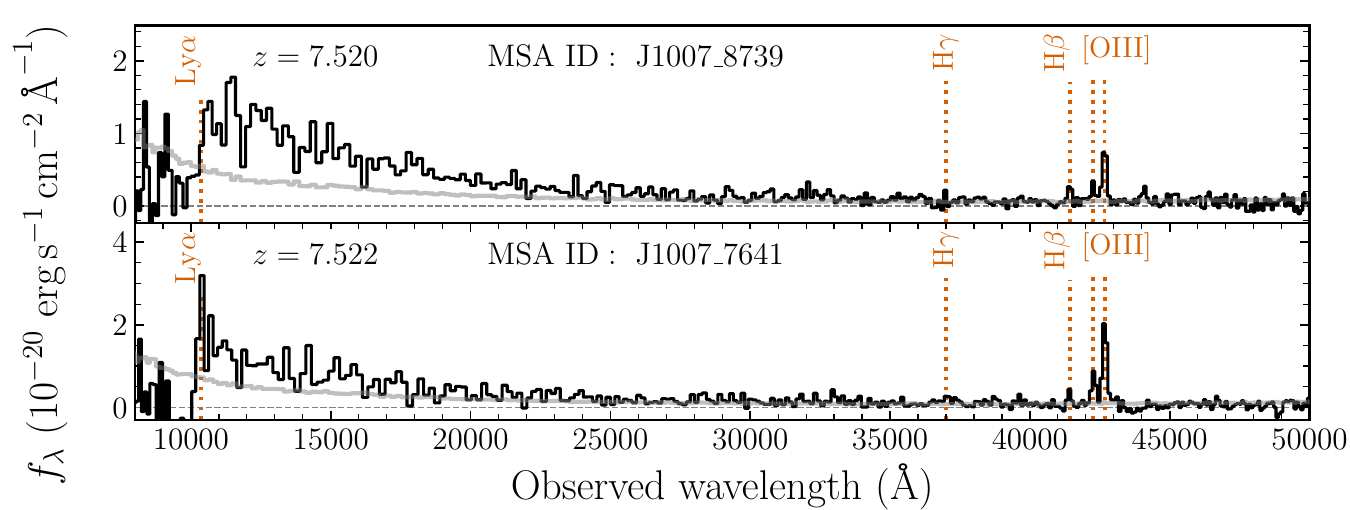}
    \caption{Discovery spectra of galaxies within $\|\Delta v_{\textrm{LOS}}\|=1500\,\rm{km}\,\rm{s}^{-1}$ to the quasar J1007$+$2115 ($z=7.5149$). The spectrum is shown in black; the vertical orange annotations highlight possible emission line features and the position of the \Lya\ break. The uncertainties ($1\sigma$) on the spectral flux are shown in gray.}
    \label{fig:J1007_cluster_gal_1D}
\end{figure*}

\begin{figure*}[ht!]
    \centering
  \includegraphics[width=0.8\linewidth]{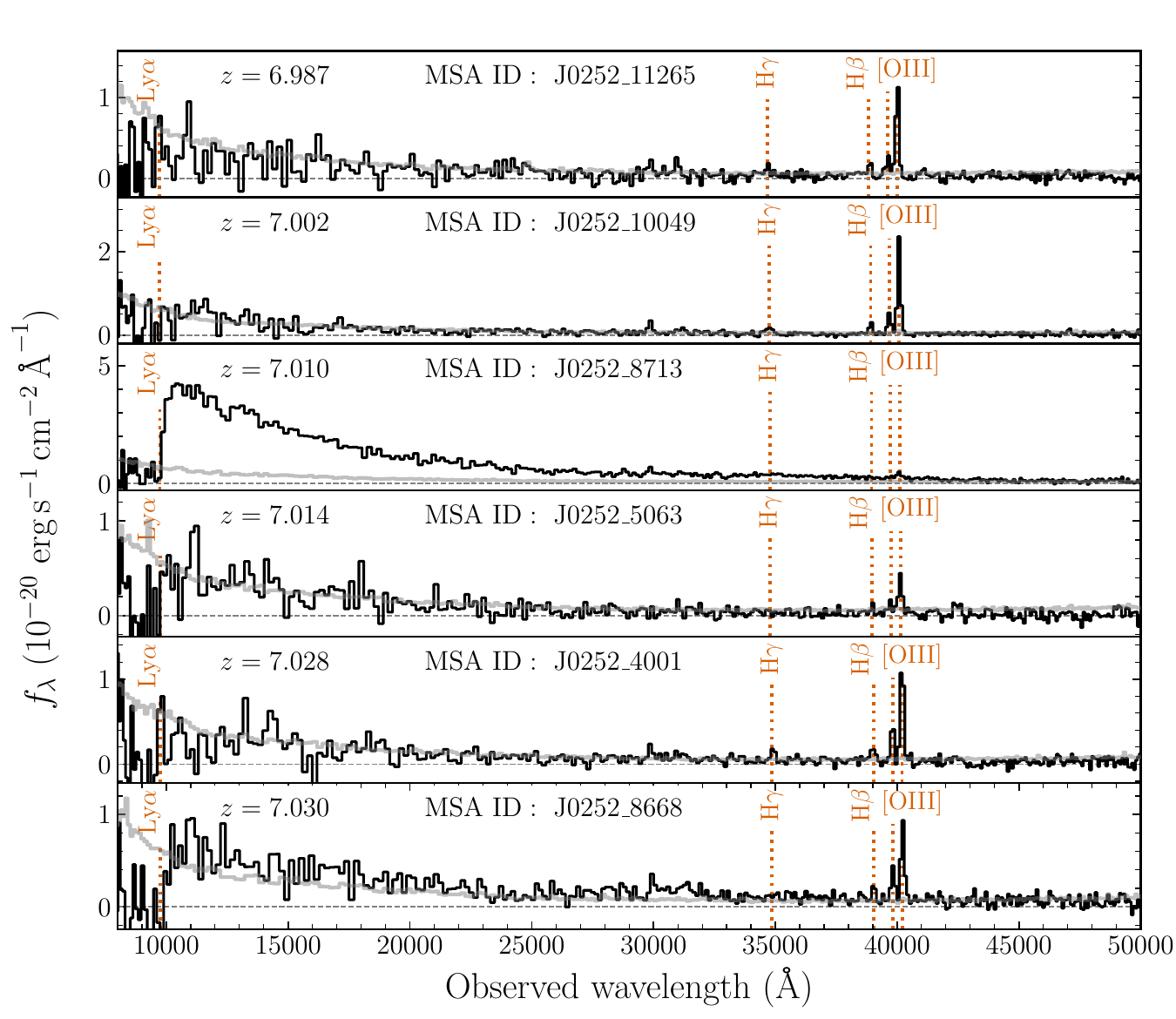}
    \caption{Discovery spectra of galaxies within $\|\Delta v_{\textrm{LOS}}\|=1500\,\rm{km}\,\rm{s}^{-1}$ to the quasar J0252$-$0503 ($z=7.00$). The spectrum is shown in black; the  vertical orange annotations highligh possible emission line features and the position of the \Lya\ break. The uncertainties ($1\sigma$) on the spectral flux are shown in gray.}
    \label{fig:J0252_cluster_gal_1D}
\end{figure*}

The NIRSpec/MSA spectroscopy used the PRISM/CLEAR disperser filter pair to provide continues spectra over $0.6\,\mu{\textrm{m}}$ to $5.3\,\mu{\textrm{m}}$ with a resolving power of $R\sim30 - 330$.
Each field was sampled with two MSA pointings using the standard 3 shutter slitlet nod pattern and were read out using the \texttt{NRSIRS2RAPID} pattern. 
We list the further observation information for each field separately: 

\begin{itemize}
    \item J1007$+$2115 field: 
To reach the desired depth, each pointing used 55 groups per integration and 2 integrations per exposure, resulting in a total exposure time of $4902\,\textrm{s}$ per pointing.
For the eMPT mask design we retained a total of 78, 64, 52, 698, 993 candidates in the first five priority classes after visual inspection. Out of these candidates we were able to assign slits to 34, 9, 9, 42, 31 candidates in these classes.
We note that 56, 35, 33, 389, and 572 of the candidates fall into the combined area of the two MSA pointings, resulting in an average targeting completeness per class of 61\% (34/56), 26\% (9/35), 27\% (9/33), 11\% (42/389), and 5\% (31/572).

\item J0252$-$0503 field:
We reduced the exposure time in this lower-redshift field to 55 groups per integration and 1 integration per exposure, resulting in a total exposure time of $2451\,\textrm{s}$ per pointing.
After visual inspection we were left with 64, 62, 96, 709, 994 candidates in the first five priority classes, of which we were able to assign 28, 11, 12, 66, 39 to slits, respectively. 
With 38, 33, 56, 385, 497 candidates falling within the union of the two MSA pointings, we calculate an average targeting completeness per class of 74\% (28/38), 33\% (11/33), 21\% (12/56), 17\% (66/385), 8\% (39/497).
\end{itemize}

We reduced the spectroscopic observations with a combination of the \texttt{CALWEBB} pipeline and the \texttt{PypeIt}\footnote{\url{https://pypeit.readthedocs.io/en/release/}} Python package \citep{Prochaska2020}. 
The J1007$+$2115 field was reduced using version 1.13.4 of the JWST Science calibration pipeline\footnote{\url{https://jwst-pipeline.readthedocs.io/en/latest}} (\texttt{CALWEBB}; CRDS context \texttt{jwst\_1215.pmap}). 
More recently, we reduced the J0252$-$0503 field with an updated pipeline (1.16.0) and CRDS context (\texttt{jwst\_1298.pmap}).
We processed the rate files with the \texttt{CALWEBB} Spec2Pipeline skipping the \texttt{bkg\_subtract}, \texttt{master\_background\_mos}, \texttt{resample\_spec}, and \texttt{extract\_1d} steps. The resulting files were then read in by \texttt{PypeIt}, which uses difference imaging for background subtraction and then co-adds the 2D spectra according to the nod pattern. 
1D spectra for all sources were optimally extracted from the coadded 2D spectra.
Flux calibration was only based on the calibration files. No additional flux calibration was carried out for the discovery spectra.

\section{Galaxy discoveries}\label{sec:galaxies}

\begin{table*}[ht!]
\footnotesize 
\centering 
\caption{Galaxy discoveries}
\begin{tabular}{llllllll} 
\small 
 Target ID &  R.A. (J2000) & Dec. (J2000) & $z_{\rm{OIII}}$  & F090W & F115W & F277W & F444W \\ 
   &  \multicolumn{2}{c}{(decimal degrees)} &   & (nJy) & (nJy) & (nJy) & (nJy) \\
\hline 
J1007\_8739 & 151.998093 & 21.263975 & 7.5201 & -16.56 $\pm$ 10.58 & 141.99 $\pm$ 11.48 & 94.30 $\pm$ 6.88 & 145.01 $\pm$ 9.43 \\ 
J1007\_7641 & 152.001669 & 21.269052 & 7.5219 & 9.16 $\pm$ 9.81 & 72.54 $\pm$ 8.70 & 72.20 $\pm$ 5.83 & 104.64 $\pm$ 7.37 \\ 
\hline
J1007\_8731 & 152.015289 & 21.264062 & 8.2672 & -2.87 $\pm$ 4.74 & 14.17 $\pm$ 4.43 & 29.60 $\pm$ 2.79 & 39.55 $\pm$ 4.04 \\ 
J1007\_13163 & 152.022878 & 21.242270 & 8.2764 & -8.38 $\pm$ 7.72 & 28.59 $\pm$ 6.92 & 39.61 $\pm$ 3.75 & 58.31 $\pm$ 5.96 \\ 
\hline 
\hline
J0252\_11265 & 43.055671 & -5.072118 & 6.9866 & 2.80 $\pm$ 7.58 & 41.09 $\pm$ 6.45 & 39.34 $\pm$ 2.59 & 134.30 $\pm$ 4.38 \\ 
J0252\_10049 & 43.067912 & -5.065894 & 7.0021 & 6.26 $\pm$ 6.22 & 29.09 $\pm$ 6.28 & 42.27 $\pm$ 3.63 & 121.92 $\pm$ 5.85 \\ 
J0252\_8713 & 43.069010 & -5.058693 & 7.0096 & 5.95 $\pm$ 51.66 & 1077.45 $\pm$ 54.73 & 1449.44 $\pm$ 34.37 & 1683.99 $\pm$ 42.55 \\ 
J0252\_5063 & 43.090385 & -5.037356 & 7.0143 & 1.64 $\pm$ 4.21 & 42.22 $\pm$ 3.85 & 27.95 $\pm$ 2.42 & 35.81 $\pm$ 3.41 \\ 
J0252\_4001 & 43.070046 & -5.032002 & 7.0275 & -6.68 $\pm$ 11.18 & 85.41 $\pm$ 12.60 & 158.18 $\pm$ 6.21 & 248.24 $\pm$ 7.77 \\ 
J0252\_8668 & 43.094657 & -5.058450 & 7.0297 & 1.93 $\pm$ 10.33 & 98.74 $\pm$ 9.14 & 181.14 $\pm$ 5.85 & 308.52 $\pm$ 8.61 \\ 
\hline
J0252\_4981 & 43.064482 & -5.037052 & 7.2717 & 4.88 $\pm$ 3.24 & 27.27 $\pm$ 3.21 & 19.82 $\pm$ 1.86 & 36.82 $\pm$ 3.10 \\ 
J0252\_8612 & 43.082074 & -5.058212 & 7.2870 & -13.06 $\pm$ 38.12 & 191.90 $\pm$ 34.27 & 154.96 $\pm$ 14.75 & 239.73 $\pm$ 22.94 \\ 
J0252\_10951 & 43.080649 & -5.070559 & 7.4036 & -29.59 $\pm$ 9.98 & 55.09 $\pm$ 9.88 & 45.87 $\pm$ 4.35 & 97.33 $\pm$ 6.52 \\ 
J0252\_10952 & 43.080611 & -5.070602 & 7.4115 & 6.59 $\pm$ 10.08 & 70.03 $\pm$ 9.91 & 42.60 $\pm$ 4.13 & 95.84 $\pm$ 6.23 \\ 
J0252\_4747 & 43.056905 & -5.035835 & 7.5178 & -1.47 $\pm$ 3.92 & 17.77 $\pm$ 3.97 & 27.00 $\pm$ 2.19 & 36.57 $\pm$ 3.29 \\ 
J0252\_6197 & 43.054927 & -5.043807 & 7.5458 & 6.09 $\pm$ 10.19 & 26.38 $\pm$ 9.67 & 56.01 $\pm$ 5.07 & 79.61 $\pm$ 6.61 \\ 
\hline 
\hline
\label{tab:galaxy_info} 
\end{tabular} 
\tablefoot{
In order to make the velocity shifts in Table\,\ref{tab:galaxy_prop} consistent with the quoted redshift here, we provide a higher accuracy for the redshift than the nominal redshift uncertainty of $\sigma_z \approx 0.001$ from the fit.}
\end{table*}

\begin{table*}[ht!]
\renewcommand{\arraystretch}{1.3}
\footnotesize 
\centering 
\caption{Galaxy properties relative to the quasars}
\begin{tabular}{ccccccccc} 
\small 
 Target ID & Priority & $\Delta v_{\rm{LOS}}$ & Angular separation & Angular separation & $M_{\rm{UV}}$  & $L_{\rm{[OIII]}5008}$ & $EW_{\rm{[OIII]}5008}$\\ 
   &  & $(\rm{km}\,\textrm{s}^{-1})$ & (arcseconds) & $(\textrm{pkpc})$ & (mag) & $(10^{42}\,\rm{erg}\,\rm{s}^{-1})$ & $(\rm{\AA})$ \\ 
\hline 
J1007\_8739 & 1 & 182 & 27.66 & 138.55 & ${-21.03}_{-0.08}^{+0.09}$& ${0.98}_{-0.10}^{+0.10}$& ${249.76}_{-35.14}^{+40.92}$\\ 
J1007\_7641 & 1 & 246 & 49.43 & 247.61 & ${-20.30}_{-0.12}^{+0.14}$& ${2.79}_{-0.14}^{+0.13}$& ${623.30}_{-60.41}^{+79.53}$\\ 
\hline
J1007\_8731 & 4 & 26486 & 78.54 & 393.40 & ${-18.67}_{-0.30}^{+0.41}$& ${0.83}_{-0.31}^{+0.09}$& ${2485.75}_{-1581.46}^{+6190.32}$\\ 
J1007\_13163 & 4 & 26811 & 116.01 & 581.08 & ${-19.44}_{-0.24}^{+0.30}$& ${0.54}_{-0.10}^{+0.10}$& ${204.88}_{-40.80}^{+48.88}$\\ 
\hline 
\hline 
J0252\_11265 & 1 & -502 & 68.47 & 358.00 & ${-19.57}_{-0.16}^{+0.19}$& ${1.95}_{-0.16}^{+0.15}$& ${545.79}_{-81.29}^{+105.26}$\\ 
J0252\_10049 & 1 & 80 & 25.94 & 135.61 & ${-19.20}_{-0.21}^{+0.26}$& ${3.45}_{-0.14}^{+0.14}$& ${862.96}_{-85.63}^{+97.40}$\\ 
J0252\_8713 & 1 & 360 & 1.26 & 6.58 & ${-23.12}_{-0.05}^{+0.06}$& ${0.41}_{-0.15}^{+0.14}$& ${21.13}_{-7.64}^{+7.63}$\\ 
J0252\_5063 & 1 & 538 & 108.05 & 564.95 & ${-19.60}_{-0.09}^{+0.10}$& ${0.56}_{-0.12}^{+0.10}$& ${639.07}_{-207.90}^{+398.92}$\\ 
J0252\_4001 & 1 & 1032 & 96.61 & 505.14 & ${-20.37}_{-0.15}^{+0.17}$& ${3.30}_{-0.18}^{+0.24}$& ${515.17}_{-66.68}^{+64.49}$\\ 
J0252\_8668 & 1 & 1114 & 90.82 & 474.85 & ${-20.53}_{-0.10}^{+0.11}$& ${2.66}_{-0.18}^{+0.17}$& ${228.44}_{-19.90}^{+25.02}$\\ 
\hline
J0252\_4981 & 1 & 10183 & 80.31 & 419.89 & ${-19.19}_{-0.12}^{+0.14}$& ${0.68}_{-0.10}^{+0.11}$& ${309.64}_{-77.89}^{+91.34}$\\ 
J0252\_8612 & 1 & 10757 & 45.74 & 239.16 & ${-21.31}_{-0.18}^{+0.21}$& ${1.58}_{-0.19}^{+0.13}$& ${1086.98}_{-376.32}^{+363.50}$\\ 
J0252\_10951 & 1 & 15124 & 58.56 & 306.18 & ${-19.98}_{-0.18}^{+0.21}$& ${1.05}_{-0.17}^{+0.13}$& ${563.18}_{-140.23}^{+183.77}$\\ 
J0252\_10952 & 1 & 15422 & 58.58 & 306.28 & ${-20.24}_{-0.14}^{+0.17}$& ${0.24}_{-0.08}^{+0.09}$& ${1902.82}_{-1267.83}^{+6965.09}$\\ 
J0252\_4747 & 2 & 19403 & 94.02 & 491.56 & ${-18.77}_{-0.22}^{+0.27}$& ${1.35}_{-0.12}^{+0.11}$& ${529.87}_{-88.94}^{+128.12}$\\ 
J0252\_6197 & 5 & 20452 & 74.79 & 391.05 & ${-19.21}_{-0.34}^{+0.50}$& ${0.46}_{-0.23}^{+0.14}$& ${1051.52}_{-730.93}^{+2010.22}$\\ 
\hline 
\hline 
\label{tab:galaxy_prop} 
\end{tabular} 
\tablefoot{
Given an accuracy on the emission line redshifts of $\sigma_z \approx 0.001$, the velocity along the line of sight has an uncertainty of $\approx 40\,\textrm{km}\,\textrm{s}^{-1}$.}
\end{table*}

We visually inspected all reduced 2D and 1D MSA spectra in the search for emission lines or a clear Lyman-$\alpha$ break to classify the sources and provide a first redshift estimate. We used the 6-band image cutouts, the photometric measurements and resulting best-fit \texttt{bagpipes} SED model as well as the photometric redshift posterior as additional information in this procedure.
The \oiiibr\,$\lambda\lambda4960.30,5008.24$ is the primary identification signature for galaxies in our targeted redshift ranges. 
To ensure a robust source classification, multiple people performed the visual inspection. 

Using the visual redshifts $z_{\rm{vis}}$ as a prior, we fit for the redshift of the \oiiibr\,$\lambda\lambda4960.30,5008.24$ line doublet in all galaxy spectra with $z_{\rm{vis}}>5$. The simple model consists of a power-law continuum and one Gaussian component for each of the doublet emission lines, coupling their redshift and FWHM. 
We summarize the coordinates, the median-fit $z_{\textrm{OIII}}$ redshift and the NIRCam fluxes of all identified galaxies within our targeted redshift ranges in Table\,\ref{tab:galaxy_info}. 
Based on the galaxy $z_{\textrm{OIII}}$ redshift and the quasar redshifts we calculate the corresponding line-of-sight velocity separation $\Delta v_{\textrm{LOS}}$ and the angular separation (both in arcseconds and in proper kpc).
From the fit to the \oiiibr\,$\lambda\lambda4960.30,5008.24$ we derive their \oiiibr\,$\lambda\lambda4960.30,5008.24$ line luminosities $L_{\rm{[OIII]}5008}$ and line equivalent width $EW_{\rm{[OIII]}5008}$.
As the F115W filter covers the rest-frame UV spectrum of the galaxies, we adopt the absolute F115W filter band magnitude as a proxy for the galaxies' absolute UV-magnitude $M_{\textrm{UV}}$.
Table\,\ref{tab:galaxy_prop} reports the galaxy properties along with their selection priority.
Uncertainties on $L_{\rm{[OIII]}5008}$ and $EW_{\rm{[OIII]}5008}$ denote the 16th to 84th posterior percentile range.

While recent clustering analyses of high-redshift quasars \citep[][]{Eilers2024, WangFeige2026} assume a line-of-sight velocity window of $\|\Delta v_{\textrm{LOS}}\|=1000\,\rm{km}\,\rm{s}^{-1}$ for their analysis, our analysis is severely limited due to low number statistics. 
Hence, we decided to use an extended line-of-sight velocity window of $\|\Delta v_{\textrm{LOS}}\|=1500\,\rm{km}\,\rm{s}^{-1}$ relative to the quasar. 
This allows us to include two galaxies in the vicinity of J0252$-$0503, whose $\|\Delta v_{\textrm{LOS}}$ extends to $\sim1100\,\rm{km}\,\rm{s}^{-1}$.
Below we briefly describe the galaxy discoveries in each individual quasar field. 

\begin{figure*}[ht!]
    \centering
    \includegraphics[width=0.48\linewidth]{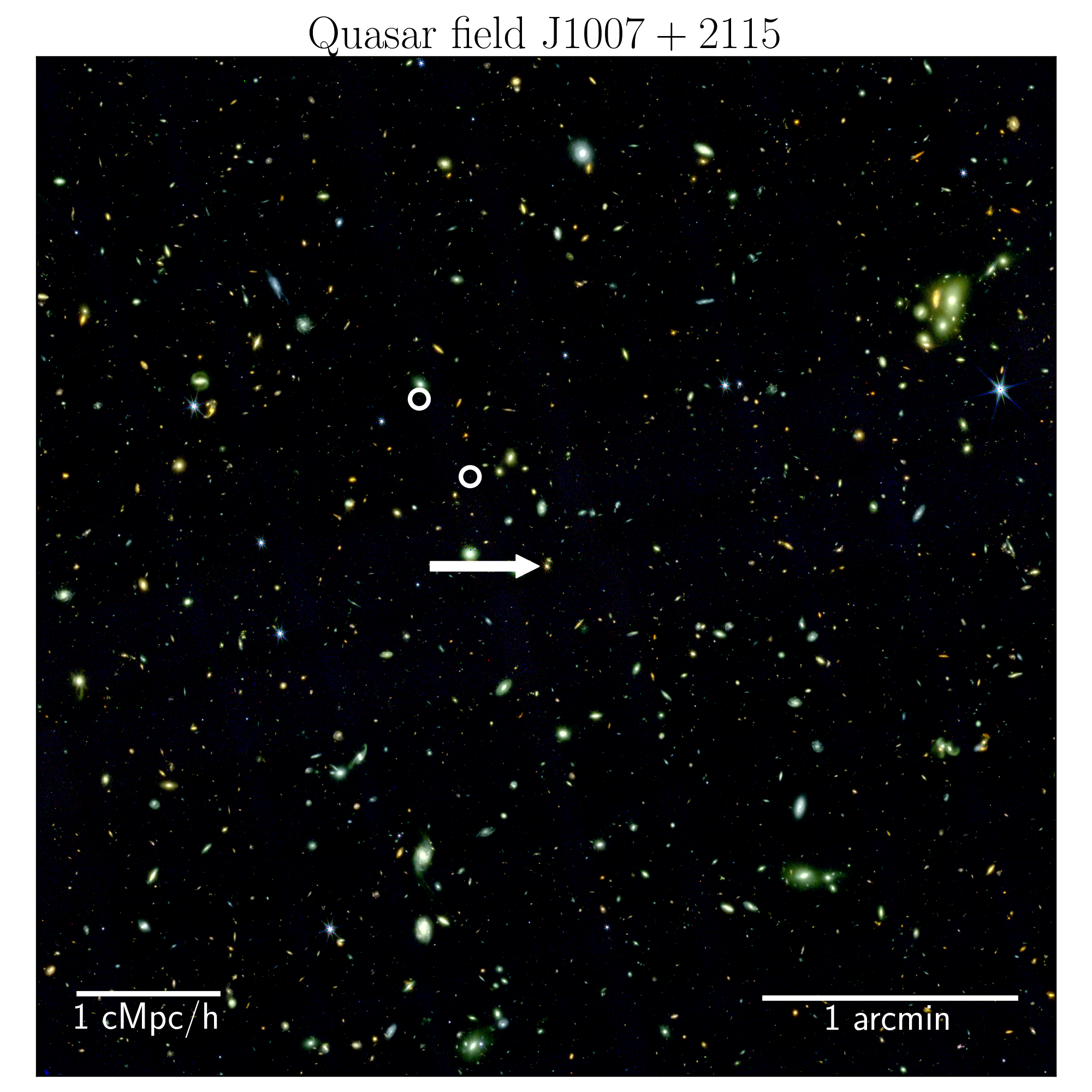}
    \includegraphics[width=0.48\linewidth]{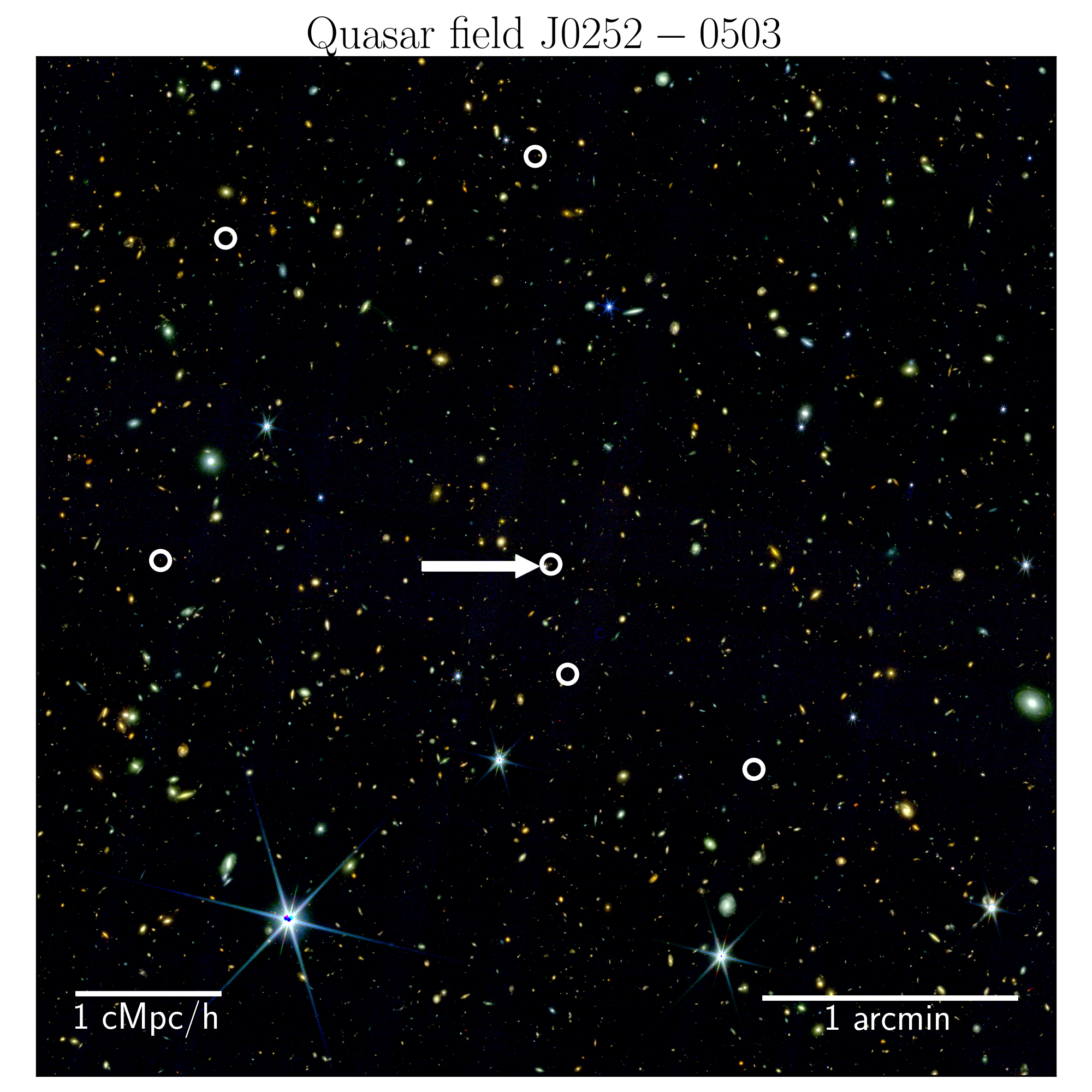}
    \includegraphics[width=0.48\linewidth]{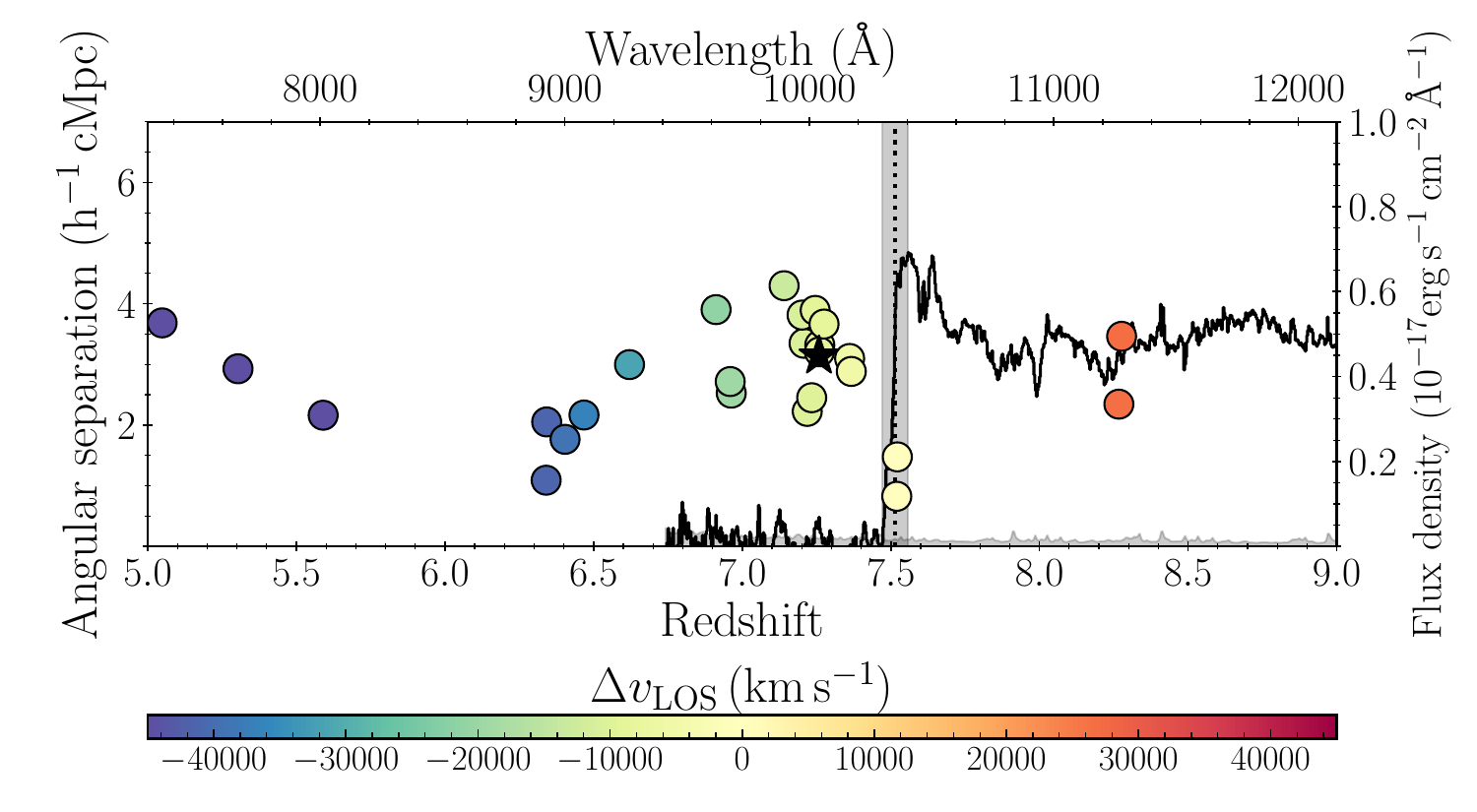}
    \includegraphics[width=0.48\linewidth]{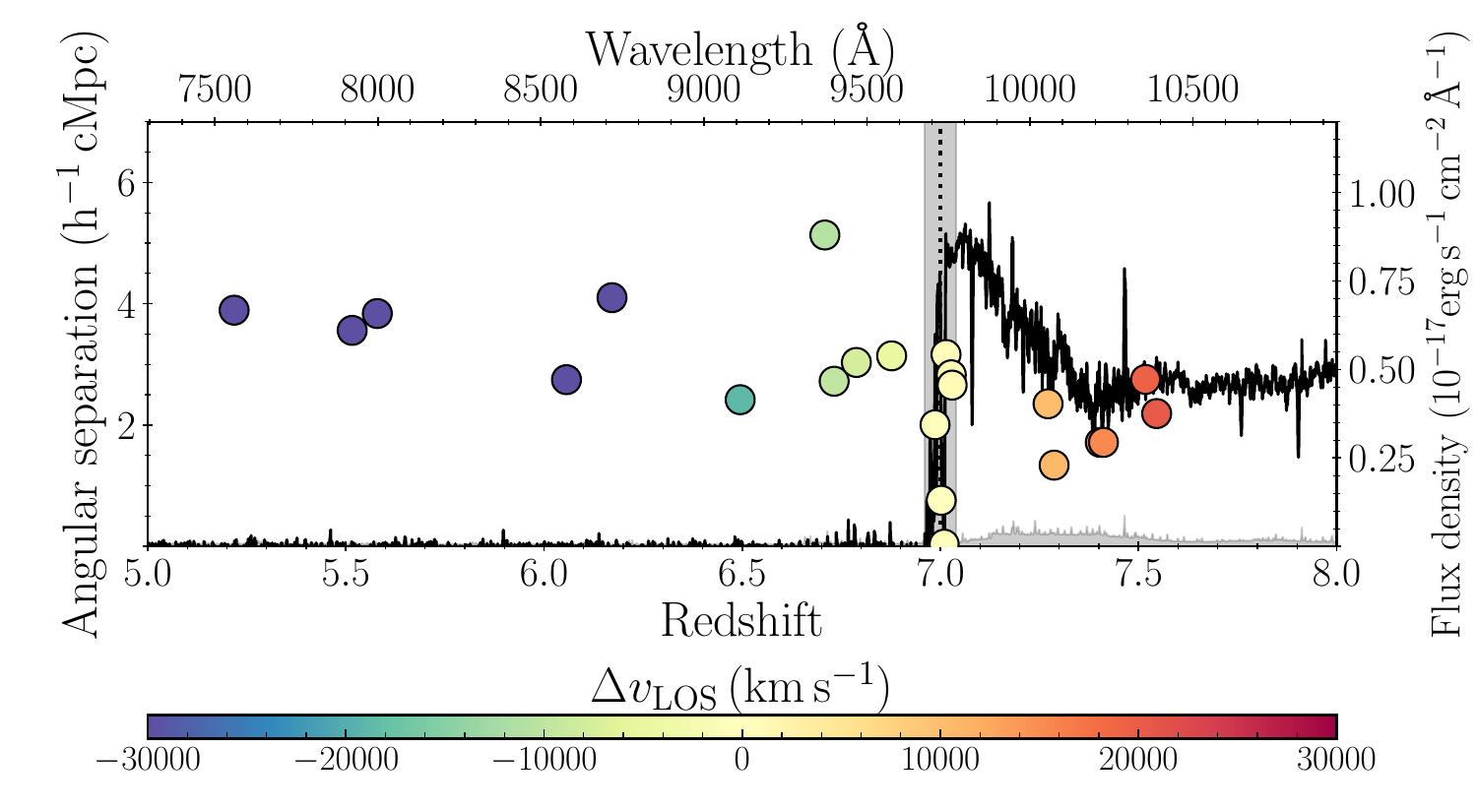}
    \caption{
    Top panels: JWST NIRCam composite image (R: F444W, G: F277W, B: F115W) of the J1007$+$2115 (left) and J0252$-$0503 (right) quasar fields. The quasar position is indicated by the tip of the white arrow. The positions of galaxies within a line-of-sight velocity window of $\|\Delta v_{\textrm{LOS}}\|=1500\,\rm{km}\,\rm{s}^{-1}$ are indicated by white circles. 
    Bottom panels: 
    Optical to near-infrared spectra (solid black lines) of the two quasars J1007$+$2115 (left) and J0252$-$0503 (right) taken from \citet{Onorato2025} with their wavelength and flux density axes on the top and right.
    We highlight the relative angular separation and redshift (left and bottom axes) of the galaxies identified in each respective field with filled colored circles.
    The quasar redshift (dotted line), coinciding with the Lyman-break in the spectrum, centers the line-of-sight velocity window used in our clustering analysis (gray shaded region). 
    The color-coding of the circles refers to the galaxies' line-of-sight velocity relative to the quasar redshift.
    The black star in the bottom left panel marks the position of a serendipitously discovered $z\approx7.3$ LRD \citep{Schindler2025}. 
    }
    \label{fig:quasar_fields}
\end{figure*}

\subsection{The J1007$+$2115 quasar field}
We discover a total of 27 $z>5$ galaxies in the field of quasar J1007$+$2115. Two galaxies lie within the clustering line-of-sight velocity range and two are found in the background of the quasar, J1007\_8731 at $z=8.27$ and J1007\_13163 at $z=8.28$. Figures\,\ref{fig:J1007_cluster_gal_1D} and \ref{fig:J1007_bg_gal_1D} show the 1D discovery spectra of the clustering and background galaxies.
In the left panels of Figure\,\ref{fig:quasar_fields} we highlight the discoveries in the J1007$+$2115 field. The top left panel 
shows the quasar position (arrow) relative to the two galaxies that will be part of the clustering analysis. 
The bottom left panel shows the angular separation of all $z>5$ discovered galaxies relative to the quasar as a function of their redshift. 
The figure highlights an overdensity of $8$ galaxies around a $z\approx7.3$ Little Red Dot (LRD; black star) that we presented in \citet{Schindler2025}, leading to a first spectroscopic measure of galaxy-LRD clustering at $z\gtrsim7$.

\subsection{The J0252$-$0503 quasar field}

\begin{figure}[h!]
    \centering
    \includegraphics[width=0.85\linewidth]{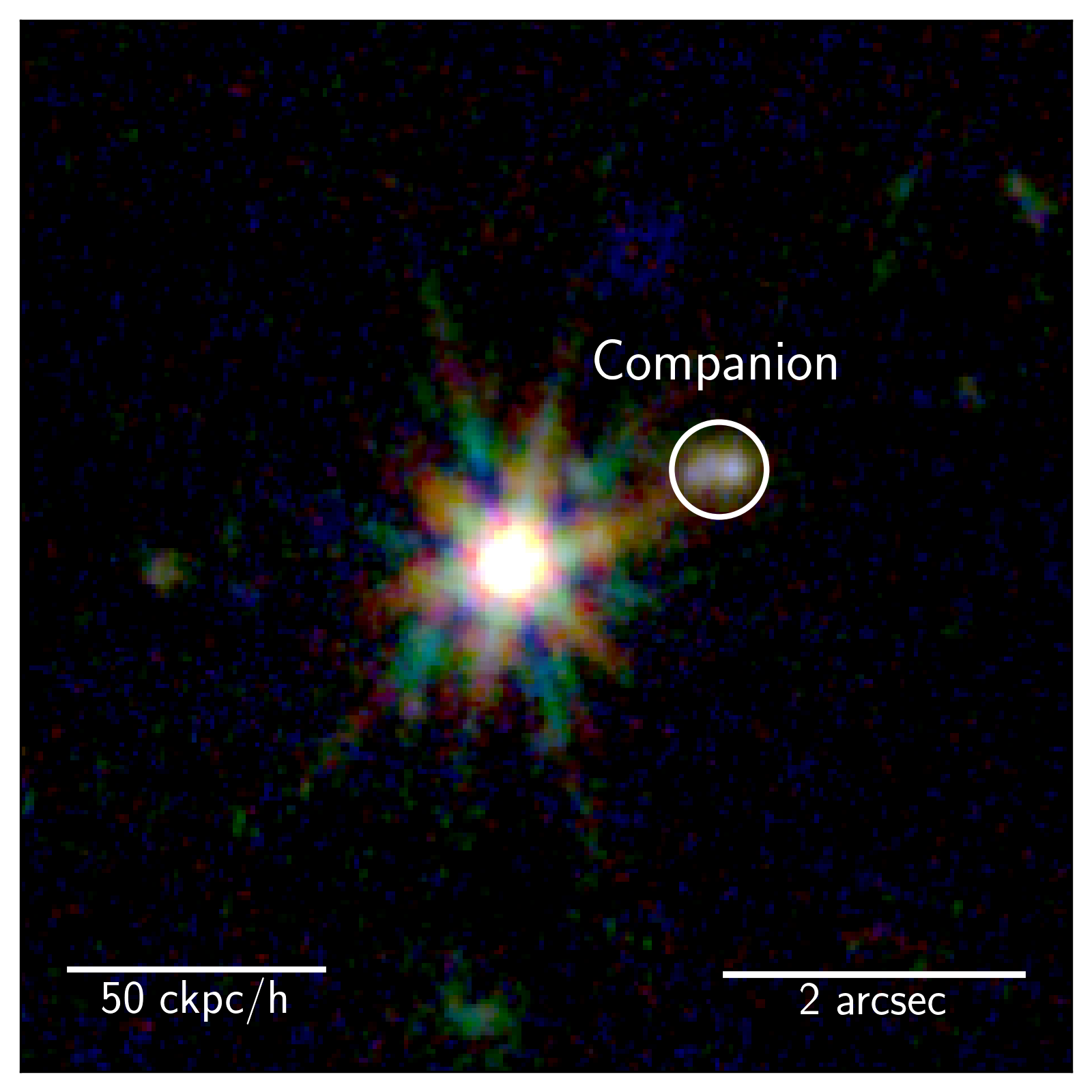}
    \caption{JWST NIRCam composite cutout ($7'' \times7''$; R: F444W, G: F277W, B: F115W) of the quasar J0252$-$0503 and its immediate environment. The quasar is the bright point source near the center. We highlight the companion galaxy, a diffuse source to the top right of the quasar, with a white circular border. The distance to the companion is $1\farcs26$, equivalent to $6.58\,\rm{pkpc}$ or $36.87\,\rm{ckpc}\,\rm{h}^{-1}$ at $z=7$.
    }
    \label{fig:J0252_companion}
\end{figure}

In the field of quasar J0252$-$0503 we identify a total of 22 $z>5$ galaxies. Of these six lie within the clustering line-of-sight velocity range and another six are found in the background of the quasar in the redshift range of $z=7.27-7.55$. We provide their 1D discovery spectra (Figures\,\ref{fig:J0252_cluster_gal_1D} and \ref{fig:J0252_bg_gal_1D}).
Both in terms of clustering and background galaxies J0252$-$0503 seems to be a richer field than J1007$+$2115. 

\subsection{A close companion to quasar J0252$-$0503}
With a transverse separation of 1\farcs26 (or $6.58\,\rm{pkpc}$) and a line-of-sight velocity of $\Delta v_{\textrm{LOS}}\approx360\,\rm{km}\,\rm{s}^{-1}$, the clustering galaxy J0252\_8713 at $z=7.01$ is in extremely close proximity to quasar J0252$-$0503. 
Figure\,\ref{fig:J0252_companion} shows a 7'' by 7'' cutout of the JWST NIRCam composite, highlighting the quasar and J0252\_8713. 
Its 1D spectrum, depicted in the third row from the top in Figure\,\ref{fig:J0252_cluster_gal_1D}, shows strong rising continuum emission toward the rest-frame UV and the weakest \oiiibr-emission lines in our sample (Table\,\ref{tab:galaxy_prop}). 
Given the proximity of quasar and galaxy, it may present one of the most distant quasar-galaxy merger systems known to date.
Future work will focus on this unique system.

\section{Quasar-galaxy clustering at $z\simeq7.3$}\label{sec:clustering}

\subsection{Completeness of the galaxy detections}\label{sec:completeness}
It is critical to take the selection function of our galaxy identifications into account when calculating the quasar-galaxy cross-correlation function. 
For our clustering analysis, we parameterize the volume using cylindrical shells. In this geometry $R$ is the radial coordinate representing the transverse comoving distance and $Z$ is the cylinder height, referring to the radial comoving distance.
Hence, we can write the effective cylindrical volume for each quasar field as 
\begin{equation}
    V_{\textrm{eff}} = \int_{Z_{\textrm{min}}}^{Z_{\textrm{max}}} \int_{R_{\textrm{min}}}^{R_{\textrm{max}}} S(R, Z) 2\pi R {\textrm{d}}R{\textrm{d}}Z \ ,
\end{equation}
where $S(R,Z)$ denotes the galaxy selection function. 
We can generally express $Z$ as a function of the Hubble parameter $H(z)$ at redshift $z$,
\begin{equation}
Z = \frac{c}{H(z)}\delta z \ ,
\end{equation}
directly connecting the galaxy redshift $z$ with the radial comoving distance $Z$. 
For our purposes we decompose the selection function into three separate, multiplicative components:
\begin{equation}
S(R, Z) = S_Z(Z) S_R(R) S_T(R) \ .
\end{equation}
Here $S_Z(Z)$ is the redshift dependent selection function, $S_R(R)$ is the radially dependent coverage selection function, and $S_T(R)$ is the radially dependent targeting selection function.
The coverage completeness $S_R(R)$ refers to the fractional radial bin area covered by the union of our NIRCam photometry and the NIRSpec/MSA follow-up. Their union is essentially the area that we can effectively select galaxy candidates from. 
However, not all of those candidates make it into MSA slits. The targeting selection function $S_T(R)$ quantifies how many of the available candidates in the MSA footprints were assigned to slits. 

We have chosen a relative line-of-sight velocity separation of $\left| \Delta v_{\textrm{LOS}}\right|\le1500\,\textrm{km}\,\textrm{s}^{-1}$ for our quasar-galaxy cross-correlation measurement. 
This translates into redshift intervals of $7.47-7.56$ (J1007$+$2115) and $6.96-7.04$ (J0252$-$0503) for the two quasar fields. 
Given that our galaxy candidate selection was tailored for the specific quasar redshifts, we assume that the selection function is a constant across the redshift ranges above. For simplicity, we conservatively set the selection function to $S_Z(Z)=1.0$ (100\%) in both cases.

We choose four radial bins to accommodate the limited number of galaxies identified in the nearby environment of the two quasars. 
The inner edge of our radial bins with a distance of $0.022\,\textrm{h}^{-1}\,\textrm{cMpc}$ ($\simeq 0.75\,\rm{arcsec}$) to the quasar was designed to include the close-by companion galaxy, J0252\_8713.
The outer edge is set to a distance of $4.47\,\textrm{h}^{-1}\,\textrm{cMpc}$ ($\simeq 150\,\rm{arcsec}$), to cover the majority of the central region of our JWST/NIRCam mosaics ($\sim 5'\times6'$). 
Between the inner and the outer edge we space the bins logarithmically, resulting in bin edges of $0.02, 0.08, 0.32, 1.19,$ and $4.47\,\textrm{h}^{-1}\,\textrm{cMpc}$ ($0.03, 0.12, 0.45, 1.70,$ and $6.38\,\textrm{cMpc}$).

\begin{figure}[ht!]
    \centering
    \includegraphics[width=\linewidth]{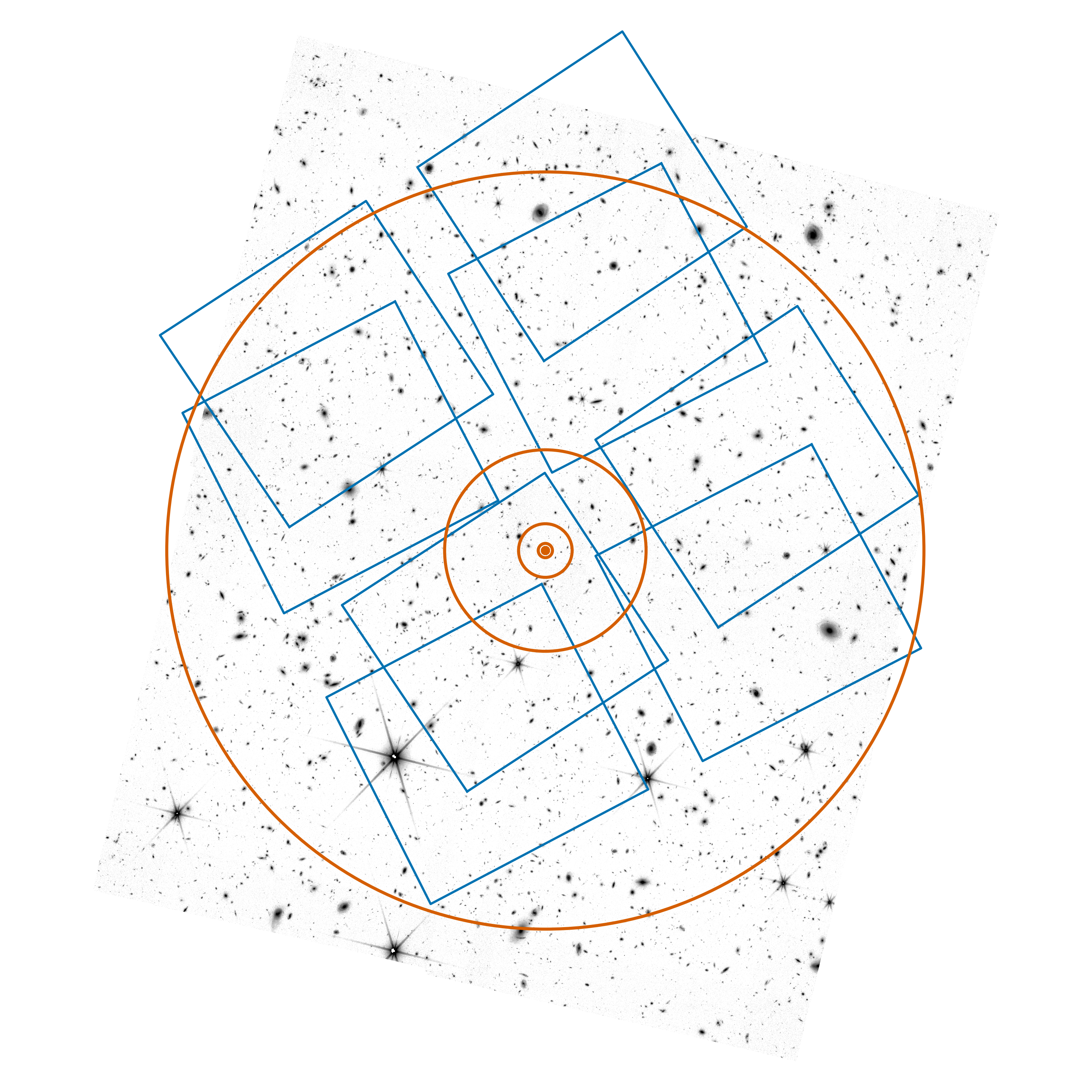}
    \caption{F277W mosaic image of the J0252$-$0503 quasar field. Overplotted are the two MSA pointings (in blue). The quasar position is depicted as an orange dot, and we highlight the radial boundaries of the four annuli with solid orange lines. This image shows that within some annuli there are gaps not covered by the NIRSpec MSA pointing and/or the NIRCam imaging that are accounted for in our coverage selection function $S_R(R)$.}
    \label{fig:coverage}
\end{figure}

The NIRCam imaging and the NIRSpec/MSA spectroscopy do not cover the full area of the radial annuli as shown in Figure\,\ref{fig:coverage}. In fact, there will be areas which have NIRCam imaging, but are not covered by the NIRSpec spectroscopic follow-up. Hence, we define the coverage selection function $S_R(R)$ as the fraction of pixels within each radial annulus that are covered by the combined NIRCam and NIRSpec observations.
As the design of the MSA masks is different for the two quasar fields, we get two sets of values for the radial coverage selection function, $S_R(R)=1.00, 1.00, 0.80, 0.79$ for quasar field J1007$+$2115 and $S_R(R)=1.00, 1.00, 0.75, 0.68$ for quasar field J0252$-$0503.

We assigned galaxy candidates to MSA slits via the eMPT tool, which introduces a targeting selection function $S_T(R)$. 
In general, the selection function introduced by object to MSA shutter assignment is nontrivial and would require forward modeling of the MSA mask design.
As we only evaluate the cross-correlation measurement in radial bins, we calculate the average of $S_T(R)$ for each radial bin. However, our prioritization of the candidates introduces dependencies on the candidate's priority class. 
Following the same approach as in \citet{Schindler2025}, we calculate a targeting selection function $S_{T}(R)$ as a function of the radial bin $R$. This targeting selection function reflects the fraction of targeted candidates to photometrically selected galaxy candidates in the available MSA area.
Due to our galaxy selection strategy this targeting selection function also depends on the candidate priority $p$. In principle, we would have to merge the $S_{T}(R)$ for different priorities within the same radial bin $R$.
However, all galaxies within the velocity range included in the clustering analysis belong to priority $p=1$, significantly simplifying the calculation.
For the J1007$+$2115 field, our MSA spectroscopy covers 0/0, 0/2, 1/2, 32/50 priority 1 galaxy candidates in the four radial bins (from small to large radii), resulting in $S_T(R)=\textrm{--}$, $0.00$, $1.00$, and $0.64$ respectively. 
In the field of quasar J0252$-$0503, the MSA spectroscopy covers 1/1, 0/0, 1/2, 25/34 priority 1 galaxy candidates in the three four bins (from small to large radii), resulting in $S_T(R)=1.00$, $\textrm{--}$, $0.5$,and $0.74$, respectively.

\subsection{Cross-correlation measurement}
In order to constrain the dark matter halo masses and duty cycles of $z\simeq7.3$ quasars, we first measure the quasar-galaxy cross-correlation function $\chi$, closely following \citet{Hennawi2006} and \citet{GarciaVergara2017}.
In a cylindrical geometry, the cross-correlation function is related to the quasar-galaxy two-point correlation function $\xi_{\textrm{QG}}$ by 
\begin{equation}
    \chi(R_{\textrm{min}}, R_{\textrm{max}}) = \frac{1}{V_{\textrm{eff}}} \int \xi_{QG}(R, Z) dV_{\textrm{eff}}\ ,
\end{equation}
where $V_{\textrm{eff}}$ is the effective comoving volume. The radial coordinate of the cylinder $R$ and the cylinder height $Z$ were introduced in the context of the selection function estimation (Section\,\ref{sec:completeness}).
We calculate the volume averaged cross-correlation $\chi(R_{\textrm{min}}, R_{\textrm{max}})$ in cylindrical shells with radii $R_{\textrm{min}}$ to $R_{\textrm{max}}$,
\begin{equation}
    \chi(R_{\textrm{min}}, R_{\textrm{max}}) = \frac{\langle QG\rangle}{\langle QR \rangle} -1 \ .
\end{equation}
The chosen radial binning scheme was introduced in section\,\ref{sec:completeness} and is also provided in Table\,\ref{tab:cross_corr}.
The cylinder height corresponds to a line-of-sight velocity difference of $\|\Delta v_{\textrm{LOS}}\|=1500\,\rm{km}\,\rm{s}^{-1}$. 
We denote the number of detected quasar-galaxy pairs in the enclosed cylindrical volumes around the two quasars by $\langle QG\rangle$, whereas $\langle QR \rangle$ refers to the number of expected galaxies over the same volumes in random (or average) regions of the Universe.
Summing over the two quasar fields we find $\langle QG\rangle$ = 1, 0, 2, and 5 quasar-galaxy pairs in the four radial bins. 

\begin{figure}[ht!]
    \centering
    \includegraphics[width=\linewidth]{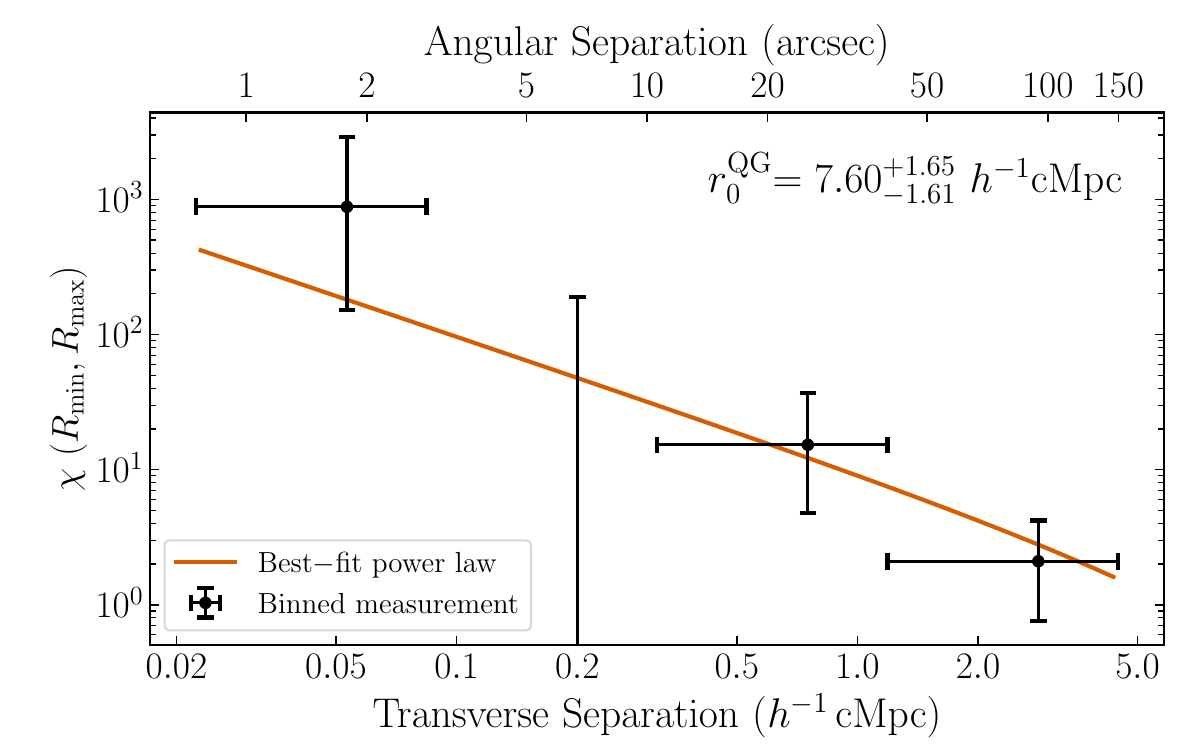}
    \caption{
    Volume-averaged cross-correlation function $\chi$ as a function of transverse separation in three radial bins (black data points).
    The orange line depicts the median posterior model of the quasar-galaxy two-point correlation function. Assuming a power law form, $\xi_{\textrm{QG}}=(r/r_0^{\textrm{QG}})^{-2.0}$, we constrain the quasar-galaxy cross-correlating length $r_0^{\rm{QG}}$.  
    The uncertainties reflect the confidence interval for a Poisson distribution for the number of clustering galaxies per bin that corresponds to $1\sigma$ in Gaussian statistics \citep{Gehrels1986}.
    }
    \label{fig:cross_corr}
\end{figure}

In order to determine the number of
random quasar-galaxy pairs $\langle QR \rangle$, we adopt an average galaxy volume density $\rho_{\textrm{gal}}$ at redshift $z$ and estimate $\langle QR \rangle$ for each quasar field and radial bin based on the effective volume of the respective cylindrical shell $V_{\textrm{eff}}$: 
\begin{equation}
\langle QR \rangle = \rho_{\textrm{gal}} V_{\textrm{eff}} \ .
\end{equation}
The volume of our observations is not large enough to determine the average galaxy volume density empirically. Hence, we decided to adopt the galaxy luminosity function of \citet{Bouwens2022} and integrate it over a UV magnitude range of $-30.0 < M_{\textrm{UV}}\le -19.2$.
The faint-end limit is set to the approximate flux limit of our galaxy selection and spectroscopic identification. 
To estimate the flux limit we applied our photometric selection criteria and the $\lambda5008.24$ emission line flux limit, at which we still successfully identified galaxies in the MSA spectroscopy, to the JAGUAR catalog \citep{Williams2018}, resulting in the approximate value of $M_{\textrm{UV}}\le -19.2$ for both fields. 
We  note that the cross-correlation measurement is extremely sensitive to $\langle QR \rangle$. In consequence, our results are directly dependent on the value of $\rho_{\textrm{gal}}$ and the choice of the faint-end integration limit.
Given our assumptions, we calculate resulting background galaxy densities of $\rho_{\textrm{gal}}=3.97\times10^5\,\textrm{Gpc}^{-3}$ and $5.90\times10^5\,\textrm{Gpc}^{-3}$, for the fields of J1007$+$2115 ($z=7.52$) and J0252$-$0503 ($z=7.00$), respectively.
We proceed to estimate $\langle QR \rangle$ for each quasar field and radial bin, taking into account the different galaxy volume densities and effective volumes. The number of random quasar-galaxy pairs for both fields are then summed for the final measurement.
We present our results in Figure\,\ref{fig:cross_corr} and Table\,\ref{tab:cross_corr}.
Integrated over all radial bins, we find that the average $z\simeq7.3$ quasar environment is overdense with a factor of  $\delta = \langle QG\rangle/\langle QR\rangle-1\approx4$. 
If we calculate the integrated overdensity for both quasar fields separately, we find a value of $\delta=4.8$ for J0252$-$0503 and $\delta=1.8$ for J1007$+$2115, reflecting the diversity of the two environments indicated by the distinct number of companion galaxies.

\begin{table}[h!]
\footnotesize 
\centering 
\caption{Quasar-galaxy cross-correlation results}
\begin{tabular}{ll|lll} 
\small 
 $R_{\rm{min}}$ & $R_{\rm{max}}$ & $\langle QG\rangle$  & $\langle QR\rangle$ & $\chi_{QG}$ \\ 
 $(\rm{cMpc}\ h^{-1})$  & $(\rm{cMpc}\ h^{-1})$ & & &  \\ 
\hline 
\multicolumn{5}{l}{$\left|\Delta v_{\rm{LOS}}\right| \le 1500\,\rm{km}\,\rm{s}^{-1}$ } \\ \hline 
0.02 & 0.08 & 1 &  0.001 & $880.73_{-729.41}^{+2027.55}$  \\ 
0.08 & 0.32 & 0 &  0.010 & $-1.00_{+0.00}^{+190.04}$  \\ 
0.32 & 1.19 & 2 &  0.123 & $15.25_{-10.49}^{+21.43}$  \\ 
1.19 & 4.47 & 5 &  1.613 & $2.10_{-1.34}^{+2.10}$  \\ 
\hline 
\multicolumn{5}{r}{$r_0^{\rm{QG}}=7.60_{-1.61}^{+1.65}\,\rm{cMpc}\ h^{-1}$} \\ 
\hline 
\hline 
\label{tab:cross_corr} 
\end{tabular} 
\end{table}

\subsection{Fitting the quasar-galaxy cross-correlation function}\label{sec:cross_corr_fit}
In order to gain further insight into the clustering properties of high-redshift quasars, we investigate the real-space quasar-galaxy two-point correlation function $\xi_{\rm{QG}}$. 
Following the literature \citep[e.g.,][]{GarciaVergara2017,Eilers2024}, we adopt a power-law parameterization,
\begin{equation}
\xi_{\textrm{QG}} = \left(r/r_0^{\textrm{QG}}\right)^{-\gamma_{\textrm{QG}}} \ , 
\end{equation}
where $r=\sqrt{R^2+Z^2}$ is the radial coordinate, $r_0^{\textrm{QG}}$ is the cross-correlation length, and $\gamma_{\textrm{QG}}$ is the power-law slope. 
Given our limited number of quasar-galaxy pairs and the resulting broad radial bins, our data cannot constrain $r_0^{\textrm{QG}}$ and $\gamma_{\textrm{QG}}$ simultaneously.  
Hence, we decide to assume a value of $\gamma_{\textrm{QG}}=2.0$ for the power law slope, to allow for comparison with the literature \citep[e.g.,][]{GarciaVergara2017, Eilers2024, HuangJiamu2026}.
In order to measure the cross-correlation length $r_0^{\textrm{QG}}$, we perform inference using a Poisson likelihood on the binned cross-correlation measurements. 
This results in a value of \rcrosscorr, which represents the median (and 16 to 84 percentile) values from our posterior.

Under the assumption that quasars and galaxies trace the same underlying dark matter distribution \citep[e.g.,][]{GarciaVergara2017}, the quasar auto-correlation function $\xi_{\rm{QQ}}$ can be expressed in terms of the galaxy-galaxy auto-correlation $\xi_{\rm{GG}}$ and the quasar-galaxy cross-correlation function $\xi_{\rm{QG}}$,
\begin{equation}
\xi_{\textrm{QQ}} = \xi_{\textrm{QG}}^2/\xi_{\textrm{GG}}. 
\end{equation}
Under the assumption that these correlation functions are power laws with the same slope, we can directly relate their correlation lengths \citep[see, e.g.,][]{ShenYue2007, Eftekharzadeh2015, GarciaVergara2017, Eilers2024, WangFeige2026}.
Assuming a value for the galaxy auto-correlation length $r_0^{\rm{GG}}$, this allows us to determine the quasar auto-correlation length $r_0^{\rm{QQ}}$ from our cross-correlation measurement. 

However, it is challenging to identify an appropriate measurement of $r_0^{\rm{GG}}$ from the literature. 
Our galaxy population was selected via the Lyman-break in JWST photometry and then confirmed spectroscopically via the detection the \oiiibr-doublet. Based on our identified population we can confirm galaxies with $M_{\textrm{UV}}\lesssim -19$ and $L_{\rm{[OIII]}5008}\gtrsim0.5\times10^{42}\,\rm{erg}\,\rm{s}^{-1}$.
This approach is unique in comparison to the literature at these redshifts, with studies either focusing on photometrically selected Lyman-break galaxies or using spectroscopically confirmed \oiiibr{} emitters.
\citet{Dalmasso2024} and \citet{Dalmasso2024a} take the first approach to determine the auto-correlation length of Lyman-break galaxies in Hubble Legacy fields and from the JWST Advanced Deep Extragalactic Survey, respectively.
Over the full GOODS regions \citet{Dalmasso2024} find an auto-correlation length $r_0^{\rm{GG}}=10.74\pm7.06\,\ h^{-1}\,\rm{cMpc}$ ($\bar{z}=7.7$; $\gamma=1.6$; $M_{\rm{UV}}<-19.8$). 
In \citet{Dalmasso2024a} the authors obtain a value of $r_0^{\rm{GG}}=3.0\pm0.7\,\ h^{-1}\,\rm{cMpc}$ ($\bar{z}=7.5$; $\gamma=1.6$) for a fainter sample, $M_{\rm{F200W}}<-15.5$. 
Apart from a slight difference in the redshift bins, the samples have distinct limiting magnitudes, resulting in the two discrepant values for $r_0^{\rm{GG}}$.

Alternatively, JWST wide field slitless spectroscopy (WFSS) is now commonly used to build large samples of \oiiibr-emitting galaxies at high redshift \citep[e.g.,][]{Matthee2023, Meyer2024, Meyer2025}.
With a general focus on quasar-galaxy clustering, two recent studies determine $r_0^{\rm{GG}}$ from \oiiibr emitters in quasar fields.
Analyzing WFSS data in six quasar fields, \citet{Eilers2024} constrain a galaxy auto-correlation length of $r_0^{\rm{GG}}=4.1\pm0.3\,\ h^{-1}\,\rm{cMpc}$ ($\bar{z}=6.25$; $\gamma=1.8$) for their identified 
\oiiibr-emitting galaxies ($L_{\rm{[OIII]}5008}>10^{42}\,\rm{erg}\,\rm{s}^{-1}$). 
Based on the 25 quasar fields of the ASPIRE survey \citep[][]{WangFeige2023}, \citet{HuangJiamu2026} find a value of $r_0^{\rm{GG}}=4.7_{-0.6}^{+0.5}\,\ h^{-1}\,\rm{cMpc}$ ($\bar{z}=6.32$; $\gamma=1.8$) for \oiiibr-emitting galaxies with $L_{\rm{[OIII]}5008}>9.4\times10^{41}\,\rm{erg}\,\rm{s}^{-1}$.
While the recent study from \citet{Shuntov2025} determines clustering properties of $z\approx7.3$ \oiiibr{} emitters from FRESCO \citep{Meyer2024}, the study does not constrain the real-space auto-correlation and $r_0^{\rm{GG}}$ in their clustering analysis. 

While these studies do not exactly reflect our selection method and criteria (redshift, UV-magnitude), in large they motivate an auto-correlation length of $r_0^{\rm{GG}}\sim5\,\ h^{-1}\,\rm{cMpc}$ for relatively bright galaxies at $z\simeq7.3$, which we adopt here.
Under this assumption we calculate a quasar auto-correlation length of \rautocorr\ for the two quasar fields.

\subsection{Dark matter halo mass and duty cycle of $z\gtrsim 7$ quasars}

In order to link the auto-correlation length of quasars to their minimum dark matter halo mass, we use predictions of the halo model framework \texttt{Halomod} \citep{Murray2013, Murray2021} using the \citet{Tinker2008} halo mass function and the \citet{Tinker2010} bias model.
We use a constant halo occupation distribution above a minimum mass threshold $M_{\textrm{halo, min}}$, equivalent to a step-function halo occupation distribution, to predict the large-scale real-space correlation function.
We tabulate the auto-correlation length and the cumulative abundance of halos $n_{\textrm{halo,min}}$ with $M>M_{\textrm{halo, min}}$ for a grid of $M_{\textrm{halo, min}}$ values and then invert the predictions to retrieve a minimum halo mass for our measured $r_0^{\rm{QQ}}$.
This results in a  minimum halo mass of \logMmin\ with a corresponding abundance of \lognhalo.

The quasar duty cycle refers to the fraction of cosmic time a galaxy spends in a quasar phase. Assuming that quasars temporarily subsample their dark matter host distribution, we can relate their number density $n_{\textrm{QSO}}$ and the abundance of halo above a certain minimum mass $n_{\textrm{halo,min}}$ to their lifetime $t_{\textrm{QSO}}$ \citep{Haiman2001, Martini2001},
\begin{equation}
n_{\textrm{QSO}} \simeq \frac{t_{\textrm{QSO}}}{t_{\textrm{H}}(z)} n_{\textrm{halo,min}} \ ,
\end{equation}
where $t_{\textrm{H}}(z)$ is the Hubble time at redshift $z$ and $t_{\textrm{QSO}}$ is the time the galaxy spends in UV-luminous
quasar phases.
The quasar duty cycle then refers to the ratio of the quasar lifetime to the Hubble time, $f_{\textrm{duty}}=t_{\textrm{QSO}}/t_{\textrm{H}}(z)= n_{\textrm{QSO}}/n_{\textrm{halo,min}}$.
We estimate the quasar number density at $z\simeq7.3$ from the quasar luminosity function \citep{Matsuoka2023}, assuming an absolute-UV magnitude range of $ -25.5 < M_{1450} \leq -30 $.
The quasar number density is very sensitive to the faint-end limit. We chose the faint-end limit to reflect that quasar J0252$-$0503 has a UV-magnitude of $M_{1450}=-25.77\,\rm{mag}$.
With the resulting quasar number density of $\log_{10}(n_{\textrm{QSO}}/\textrm{cGpc}^{-3})=-0.53$ and our inferred cumulative halo abundance $n_{\textrm{halo,min}}$, we calculate a duty cycle of \fD, resulting in a $z\simeq7.3$ quasar lifetime of \tQ. 
In order to propagate the uncertainties of the quasar luminosity function, we estimate an uncertainty of $\sigma(\log_{10}(n_{\textrm{QSO}}/\textrm{cGpc}^{-3}))\approx0.8$, adopting the average uncertainty on the luminosity function normalization, and sample the quasar density using a log-normal distribution for $n_{\textrm{QSO}}$. 
We combine these with realizations drawn from our best-fit posterior for $n_{\textrm{halo,min}}$ and report the 16th to 84th percentiles as uncertainties.
Due to the exponential cutoff in the halo mass function, the duty cycle is very sensitive to the minimum dark matter halo mass. 
As a consequence, the uncertainties on the inferred duty cycle are dominated by our uncertainties in the minimum dark matter halo mass estimate rather than the uncertainties on the quasar number density.

\section{Discussion}\label{sec:discussion}

\subsection{Robustness of the result} 
The small number of clustering galaxies identified (8) in a sample of just two quasars, results in large statistical uncertainties.
In addition, the reported field-to-field variation in the number of quasar companion galaxies \citep[][]{Eilers2024, WangFeige2026} underline the effect of cosmic variance on small samples. 
Consequently, our statistical measurement uncertainties are likely underestimated.

Furthermore, our analysis carries implicit and explicit assumptions, that may introduce systematic biases on our results. 
In the following, we quantify the systematic effects on our main results due to two of our main assumptions, the background galaxy number density and the galaxy auto-correlation length.

To calculate the volume averaged cross-correlation function, we needed to calculate the expected number of random quasar-galaxies pairs. We used the \citet{Bouwens2022} galaxy luminosity function to estimate the number of galaxies expected in our effective cylindrical shells. 
In this calculation, the value of faint-end integration limit, $M_{\rm{UV,faint}}=-19.2$, has a significant impact on the results. 
Choosing a fainter limit of $M_{\rm{UV,faint}}=-19.1$ naturally results in a lower cross-correlation length, $r_{0}^{\textrm{QG}}\approx7.0\,\textrm{h}^{-1}\,\textrm{cMpc}$, a lower auto-correlation length, $r_0^{\textrm{QQ}}\approx10.8\,\textrm{h}^{-1}\,\textrm{cMpc}$, and a lower minimum dark halo mass, $\log_{10}(M_{\textrm{halo, min}}/\textrm{M}_{\odot})= 11.4$.
Accordingly, a brighter limit of $M_{\rm{UV,faint}}=-19.3$ has the opposite effect ($r_{0}^{\textrm{QG}}\approx8.24\,\textrm{h}^{-1}\,\textrm{cMpc}$, $r_0^{\textrm{QQ}}\approx14.8\,\textrm{h}^{-1}\,\textrm{cMpc}$, $\log_{10}(M_{\textrm{halo, min}}/\textrm{M}_{\odot})= 11.8$).

In order to infer the quasar auto-correlation length we needed to adopt a galaxy auto-correlation length measurement. We discussed the challenge of identifying an appropriate measurement in the literature (Section\,\ref{sec:cross_corr_fit}).
In the end we chose to adopt a value of $r_0^{\rm{GG}}=5\,\ h^{-1}\,\rm{cMpc}$ for this purpose. 
Varying the galaxy auto-correlation length by $\Delta r_0^{\rm{GG}}/(\ h^{-1}\,\rm{cMpc})=+1(-1)$ results in changes of $\Delta r_0^{\textrm{QQ}}/(\ h^{-1}\,\rm{cMpc})\approx-2.1(+3.2)$ for the quasar auto-correlation length. These translate into minimum dark matter halo mass differences of $\Delta \log_{10}(M_{\textrm{halo, min}}/\textrm{M}_{\odot})= -0.25/+0.32$.

The systematic biases introduced by the galaxy background density and the galaxy auto-correlation length (over their respective value ranges) are about two times smaller than our current statistical uncertainties. 
While the assumptions lead to notable systematic uncertainties, we argue that the increase of the sample size, which reduces statistical uncertainties and the impact of cosmic variance, should be the first priority of future quasar-galaxy clustering studies at $z\gtrsim7$.
However, we would like to highlight that, at the time of writing, there are only eight quasars known at $z>7$ \citep{Fan2023}, and thus our study represents one-quarter of the available sample size. 

\begin{figure}[ht!]
    \centering
    \includegraphics[width=\linewidth]{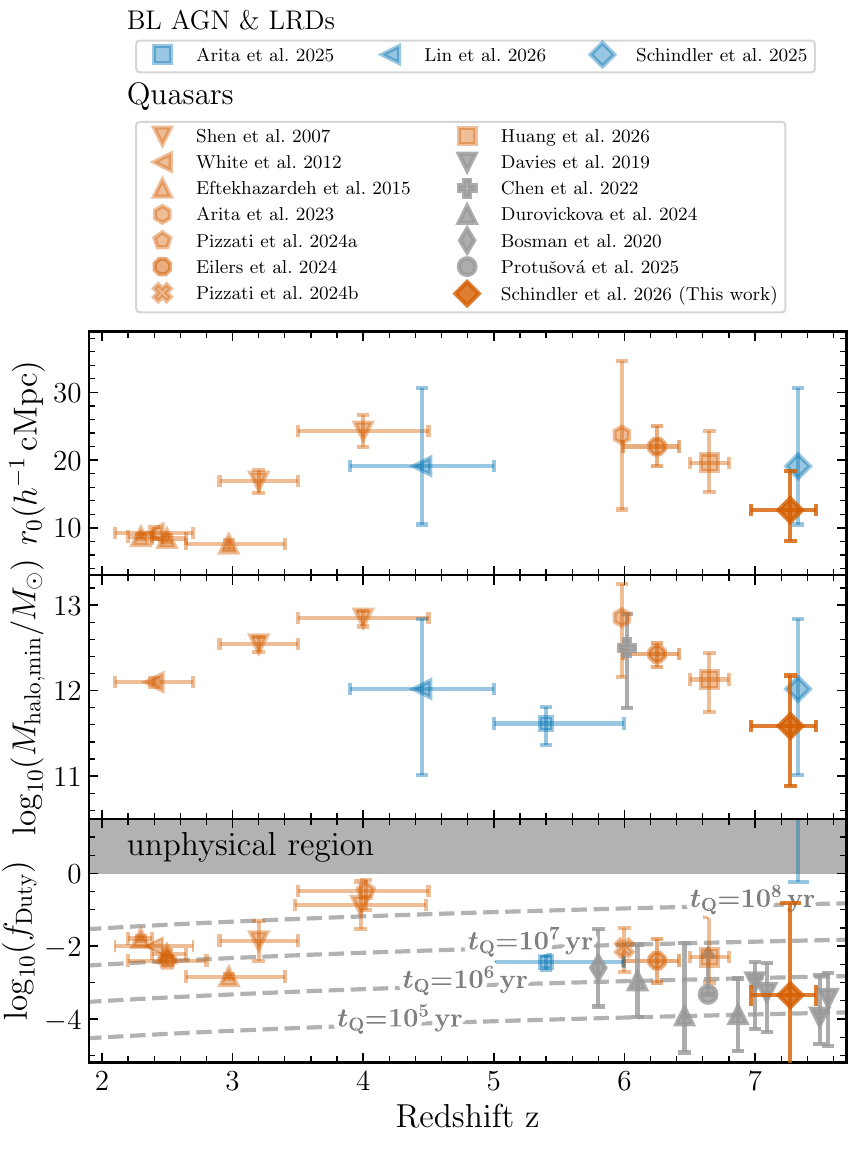}
    \caption{Redshift evolution of the auto-correlation length, the minimum dark matter halo mass, and the duty cycles of UV-luminous quasars (from top to bottom).
    The results from the quasar auto- or cross-correlation studies are highlighted in orange \citep[][]{ShenYue2007, White2012, Eftekharzadeh2015, Arita2023, Pizzati2024a, Eilers2024, Pizzati2024, HuangJiamu2026}, whereas the gray symbols refer to measurements from quasar proximity zones \citep[gray cross;][]{ChenHuanqing2022} and/or IGM damping wings \citep{Davies2019, Durovcikova2024}.
    The blue symbols indicate the results related to faint broad-line AGN \citep{Arita2025, Lin2026a} and LRDs \citep{Schindler2025}.
    }
    \label{fig:clustering}
\end{figure}

\subsection{Host dark matter halo masses of UV-luminous quasars}

Linking quasars (or AGN) to their host dark matter halos offers critical insight into the formation histories of these systems in the context of large-scale structure evolution.
\citet{Arita2023} measured the auto-correlation function of a sample of 107 quasars over an area of $891\,\rm{deg}^2$. The quasar sample was largely comprised of sources identified by the Subaru High-z Exploration of Low Luminosity Quasars \citep[SHELLQs;][]{Matsuoka2016, Matsuoka2022} with fainter UV-magnitudes ($M_{1450}\geq-25\,\rm{mag}$) at $z\sim6$. 
At scales of $10-1000\,h^{-1}\rm{cMpc}$ the authors measure an auto-correlation length of $r_0^{QQ}=23.7\pm11$ (assuming a power-law model \citep[cf.][]{Pizzati2024}) and use \texttt{Halomod} to infer a dark matter halo mass of $\log_{10}(M_{\textrm{halo}}/\textrm{M}_{\odot})=12.9_{-0.7}^{+0.4}$.

Enabled by the wide-field slitless spectroscopy available with the JWST/NIRCam instrument two recent studies constrain the quasar-galaxy cross-correlation function at $z\approx6.3$ \citep{Eilers2024} and $z\approx6.5$ \citep{HuangJiamu2026, WangFeige2026}. 
Both studies identify \oiiibr-emitting galaxies at distances of $<10\,h^{-1}\rm{cMpc}$ around high-redshift quasars without any photometric pre-selection.
In \citet{Eilers2024}, the authors investigate the fields of four bright ($M_{1450}\sim-27.5\,\rm{mag}$) $z\sim6.3$ quasars, finding a quasar auto-correlation length of $r_0^{QQ}=22.0_{-2.9}^{+3.0}$ and a host dark matter halo mass of 
$\log_{10}(M_{\textrm{halo, min}}/\textrm{M}_{\odot})=12.43_{-0.15}^{+0.13}$.
More recently, the ASPIRE program \citep{WangFeige2023} provided measurements of quasar-galaxy clustering at $z\sim6.5-6.8$ based on 25 quasar fields. These quasars span a range of absolute magnitudes, $M_{1450}\approx-27.5$ to $-25\,\rm{mag}$ with a median of $M_{1450}=-25.91$.
Using the ASPIRE sample \citet{HuangJiamu2026} measure a quasar auto-correlation length of $r_0^{QQ}=19.6^{+4.7}_{-4.3}$  
and infer a host dark matter halo mass of 
$\log_{10}(M_{\textrm{halo, min}}/\textrm{M}_{\odot})=12.13^{+0.31}_{-0.38}$.
It is important to underline that both \citet{Eilers2024} and \citet{WangFeige2026} find a large scatter in the number of detected companion galaxies, indicating significant cosmic variance in quasar environments and possibly a wide distribution of halo masses.

By measuring the density field in quasar proximity zones for ten high-resolution $z\sim6$ quasar spectra and comparing the observed density cumulative distribution function with simulations, \citet{ChenHuanqing2022}  infer a typical host dark matter halo mass of $\log_{10}(M_{\textrm{halo, min}}/\textrm{M}_{\odot})=12.5_{-0.27}^{+0.4}$ for bright ($M_{1450}<-26.5\,\rm{mag}$) quasars.
Their result is in excellent agreement with the other studies of quasar clustering at $z\sim6$ discussed above \citep{Arita2023, Eilers2024}.

Figure\,\ref{fig:clustering} provides a visual overview over the discussed literature results by depicting the auto-correlation length and the minimum dark matter halo mass as a function of redshift in the top two panels.
Taken in context with lower redshift measurements \citep{ShenYue2007, White2012, Eftekharzadeh2015}, the $z\gtrsim6$ results paint a coherent picture in which quasars are hosted in massive, $\log_{10}(M_{\textrm{halo, min}}/\textrm{M}_{\odot})\gtrsim12$ dark matter halos over much of cosmic time ($z=2-7$).

Extending these investigations to $z\simeq7.3$, we find a lower auto-correlation length of \rautocorr\ and dark matter halo mass \logMmin\ than the studies at $z=6-7$.
Adding our result to the $z>6$ literature (EIGER, ASPIRE), a decline of the quasar host dark matter halo mass emerges at the highest redshifts, presenting a first sign of nonmonotonic clustering evolution. 
We do not think that this decline is due to a luminosity dependence of quasar clustering. 
Even though the two quasars in our work are much less luminous than the EIGER quasars, they have close to the same mean luminosity as the ASPIRE sample.
Additionally, quasar clustering at $z\lesssim4$ has not been observed to depend on luminosity \citep{Porciani2006,ShenYue2009,HeWanqiu2018}.
Considering the buildup of large scale structures in the early Universe, one might naturally expect the typical quasar host $M_{\textrm{halo, min}}$ to decrease at the highest redshifts.
On the other hand, we caution that the large cosmic variance discovered in quasar environments \citep[][]{Eilers2024, WangFeige2026} might significantly skew our results based on only two quasar fields.

\subsection{The quasar duty cycle and implications for SMBH growth}
In the standard model of SMBH growth \citep{Salpeter1964} SMBH mass accumulates exponentially with a timescale of $t_{\rm{S}}=45\,\rm{Myr} \ (\epsilon/0.1)(L_{\rm{bol}}/L_{\rm{Edd}})^{-1}$, 
where $\epsilon$ is the radiative efficiency of the accretion process. 
The standard scenario assumes that SMBH growth proceeds continuously, resulting in simple exponential light curves. 
Consequently, these massive SMBHs will appear as luminous quasars for most of their cosmic history.
However, in this framework there is not enough time to build the $10^9\,\rm{M}_\odot$ SMBHs observed in $z\sim7.5$ quasars from $\sim100\,\rm{M}_\odot$ stellar remnants \citep[see, e.g.,][for a recent review]{Inayoshi2020}. 
This challenge could be overcome by continuous Eddington-limited accretion onto more massive seeds with $>1000\,\rm{M}_\odot$ that may form via exotic processes \citep[e.g.,][]{Omukai2001, Oh2002, Bromm2003, Omukai2008, Devecchi2009}. 
Alternatively, a faster mode of SMBH growth \citep[e.g.,][]{Begelman1979, Volonteri2005, Ohsuga2005, Madau2014}, with a much smaller Salpeter time $t_{\rm{S}}$, would relieve the tension in the standard scenario. 

However, this discussion ignores that SMBH growth can occur in intermittent phases, characterized by the quasar duty cycle $f_{\rm{duty}}=t_{\textrm{QSO}}/t_{\textrm{H}}(z)$ or the  quasar lifetime $t_{\textrm{QSO}}$.
High quasar duty cycles ($f_{\rm{duty}}\sim1$), reflecting quasar lifetimes on the order of the Salpeter time, have been postulated to explain the rapid growth of SMBHs in the early Universe \citep{Haiman2001a,Martini2004, Volonteri2012}.
At redshifts of $z\sim4$ clustering measurements suggest that SMBH growth occurs in long quasar phases with duty cycles of $f_{\rm{duty}}\sim33\%$, corresponding to $t_{\rm{Q}}\sim500\,\rm{Myr}$ \citep{ShenYue2007, Pizzati2024a}.
This situation seems to change at $z\gtrsim6$, where recent clustering measurements \citep[][]{Eilers2024, Pizzati2024, HuangJiamu2026} point toward duty cycles of $\simeq1\%$ equivalent to quasar lifetimes of $t_{\rm{Q}}<10\,\rm{Myr}$.
Based on our quasar-galaxy clustering measurements, we find a quasar lifetime of $t_{\rm{Q}}\sim0.31\,\rm{Myr}$ at $z\simeq7.3$ ($f_{\rm{duty}}\sim0.04\%$). 
The bottom panel in Figure\,\ref{fig:clustering} depicts the quasar duty cycle as a function of redshift and shows our result in comparison with the literature. We note the significant statistical uncertainties arising from uncertainties on the quasar luminosity function \citep{Matsuoka2023} and the exponential cutoff of the halo mass function.

Interestingly, the short quasar lifetime inferred from quasar-galaxy clustering at $z=6-7$ and our results are consistent with independent constraints on the UV-luminous duty cycle from proximity zones \citep[e.g.,][]{Eilers2017, Andika2020, Morey2021, Satyavolu2023} and quasar damping wings \citep[e.g.,][]{Davies2019, Durovcikova2024}.
These studies analyze the observed flux transmission profiles in quasar spectra, which are sensitive to the number of hydrogen ionizing photons emitted during quasar phases and the neutral fraction of the surrounding intergalactic medium (IGM). 
Modeling the observed transmission profiles, then allows inferring the quasar lifetime for individual sources. 

These short timescales for quasar activity at $z\gtrsim6$ present an additional challenge to the formation and early growth of SMBHs in the early Universe. 
On the one hand, this result could point toward a change in accretion physics at high redshift toward a radiatively inefficient mode of accretion. A radiative efficiency much smaller than the canonical value for thin accretion disks $\epsilon=0.1$, would naturally result in Eddington-limited accretion with rapid super-critical mass accretion rates and Salpeter times $t_{\rm{S}}$ much smaller than $45\,\rm{Myr}$.
In this scenario stellar remnants from Pop\,III stars would grow rapidly enough to explain the existence of billion solar mass SMBHs in the $z>7$ quasar population.
Hence, there would be no need for massive initial seeds with ($>1000\,\rm{M}_\odot$) formed via exotic processes. 

On the other hand, long periods of SMBH growth at $z\gtrsim6$ could occur in UV-obscured phases. 
Several simulations suggest that a large fraction of $z\gtrsim6$ SMBH growth may be obscured \citep{Trebitsch2019, Ni2020, Lupi2022, Vito2022,Bennett2024}. 
With dense, high column density gas prevalent in the inner regions of high-redshift galaxies, a large fraction of SMBH may remain highly obscured \citep[e.g., $f_{\rm{obs}}>99\%$ at $z\ge7$;][]{Ni2020}. Only the onset of strong quasar feedback is able to clear a small fraction of available sight lines with the majority ($f_{\rm{obs}}\approx90\%$) still being highly obscured ($N_{\rm{H}}>10^{23}\,\rm{cm}^{-2}$).

Studies of infrared-selected AGN/quasars indicate that the obscured:unobscured fraction reaches 1:1 at $z\gtrsim2$ \citep[e.g.,][]{Glikman2018, Lacy2020}. 
This work underlines that it is the fraction of moderately obscured objects that seem prevalent at high luminosities and high redshifts.
Observations at $z\gtrsim3$ \citep{Vito2018, Circosta2019, DAmato2020} provide evidence for dense gas in the inner regions of AGN/quasar host galaxies responsible for significant obscuration. Based on observations of the interstellar medium in galaxies and quasars, \citet{Gilli2022} conclude that 80-90\% of SMBH growth at $>6$ will be obscured.
Additional observational evidence for a high obscured fraction is presented in \citet{Endsley2023}, who report the discovery of an obscured radio-loud AGN at $z\approx6.9$. If accreting at the Eddington limit, the source would host a $1.6\times10^{9}\,\rm{M}_\odot$ SMBH with an intrinsic bolometric luminosity of $5\times10^{13}\,\rm{L}_\odot$.
Given the expected sky density of $z\sim7$ quasars of $\sim0.001\,\rm{deg}^{-2}$ \citep{Wang2019}, the discovery of this source in a field of  $1.5\,\rm{deg}^2$ suggests a large obscured fraction at $z\gtrsim7$.
Taken at face value, the constraints on the quasar duty cycle would result in $100-1000$ obscured quasars for every unobscured one. This implies a large population of obscured quasars at $z\gtrsim6$.
However, the search for obscured quasars, even up to cosmic noon, entails considerable observational challenges \citep[e.g.,][]{Ishikawa2023,WangBen2025}.
Recently, \citet{Matsuoka2025} followed-up  UV-selected sources with strong \Lya\ emission with JWST/NIRSpec.
The authors find 7 out of their 11 targets to exhibit broad Balmer emission lines and classify them as obscured quasars.
Based on this sample, the authors argue that there is a significant obscured fraction at $z>6$ with number densities similar to their unobscured counterparts. 
Yet, these results are still far away from the $100-1000:1$ obscured:unobscured ratio implied by the short quasar duty cycles.

At first glance, abundant broad-line AGN appearing as compact red sources in JWST/NIRCam imaging---called little red dots (LRDs)---may appear as intriguing candidates for obscured SMBH growth \citep{Kocevski2023, Harikane2023, Matthee2024, Greene2024, Maiolino2024b}. 
Initially interpreted as moderately obscured AGN superimposed on a galaxy stellar component or a fraction of unattenuated scattered AGN light \citep{Harikane2023, Matthee2024, Greene2024}, their intrinsic luminosities and SMBH masses would rival faint quasars. In this scenario, their number densities place them factors of 100-1000 \citep{Matthee2024, Greene2024, Kocevski2025, Kokorev2024} above the bolometric quasar luminosity function \citep{ShenXuejian2020}.
We note, however, that recent results argue for a much lower bolometric luminosity corrections \citep{Greene2026}.

Clustering can relate LRDs to their host dark matter halo population to understand if they could serve as obscured counterparts to UV-luminous quasars. 
\citet{Arita2025} and \citet{Lin2026a} probe the clustering properties of low-luminosity AGN, including a mix of LRDs and normal broad-line AGN, in their samples. 
Both studies find smaller host dark matter halo masses compared to UV-luminous quasars (see blue symbols in Figure\,\ref{fig:clustering}) at their respective redshifts.
A dependence of clustering properties on luminosity (or SMBH mass) could play a role, given that these broad-line AGN identified with JWST are much fainter ($L_{\textrm{bol}}\sim10^{44}\,\rm{erg}\,\rm{s}^{-1}$) than the quasar population ($L_{\textrm{bol}}\sim10^{46}\,\rm{erg}\,\rm{s}^{-1}$). 
Additionally, these samples contain a mixture of type-1 AGN with broad lines and LRDs and may represent a mixed source population.
In contrast, the discovery of an overdensity of galaxies around one LRD at $z\approx7.3$ \citep{Schindler2025} implies a larger dark matter halo ($\log_{10}(M_{\textrm{halo, min}}/\textrm{M}_{\odot})\approx 12.0_{-1.0}^{+0.8}$) mass than our measurement for UV-luminous quasars at the same redshift (\logMmin). 
Given the abundant nature of LRDs, this would imply unphysically high ($>1$) LRD duty cycles. 
However, given the significant statistical and systematic uncertainties (e.g., cosmic variance), it is likely that this system is an outlier of the general LRD population  \citep[see discussions in][]{Pizzati2025, Schindler2025}.
However, it underlines that the host dark matter masses of LRDs can extend to the high-mass end, implying high duty cycles.
This result at least establishes the plausibility of LRDs as an obscured precursor population to UV-luminous quasars. 

\section{Summary}\label{sec:summary}

In this paper, we present the results of the JWST Cycle\,1 program GO 2073 “Towards Tomographic Mapping of Reionization Epoch Quasar Light-Echoes with JWST” (PI: J. Hennawi). The program was designed to identify galaxies in the surrounding and background of two high-redshift quasars, J1007$+$2115 at $z=7.51$ and J0252$-$0503 at $z=7.00$. 
While we aim to utilize these galaxy discoveries to tomographically map the ionized regions around those quasars in the future, our present work fulfills the direct goal of the proposal, providing a first constraint on quasar clustering at $z\simeq7.3$.
We summarize the main findings below:

\begin{itemize}
    \item In the J1007$+$2115 quasar field, we discover a total of 28 $z>5$ galaxies, of which two are within a velocity window of $\|\Delta v_{\textrm{LOS}}\|=1500\,\rm{km}\,\rm{s}^{-1}$ relative to the quasar and another two are in the background ($z=8.27$ and $z=8.28$).
    Furthermore, we note that this quasar field holds an overdensity associated with an LRD, as presented in \citet{Schindler2025}.
    In the J0252$-$0503 quasar field we identify a total of 23 $z>5$ galaxies, of which six lie within a velocity window of $\|\Delta v_{\textrm{LOS}}\|=1500\,\rm{km}\,\rm{s}^{-1}$ to the quasar and six are found in the background ($z=7.27-7.55$).
    \item We  particularly highlight the galaxy J0252\_8713 at $z=7.01$ ($\Delta v_{\textrm{LOS}}=-502\,\rm{km}\,\rm{s}^{-1}$) at an angular separation of 1\farcs26 ($6.58\,\rm{pkpc}$ or $37\,\rm{h}^{-1}\rm{ckpc}$, also see Figure\,\ref{fig:J0252_companion}). J0252\_8713, presents an exciting opportunity to further study quasar-galaxy interactions and the role of mergers in the activation and fueling of quasar phases. 
    \item Based on the galaxies within $\|\Delta v_{\textrm{LOS}}\|=1500\,\rm{km}\,\rm{s}^{-1}$ to the two quasars, we measure the quasar-galaxy cross-correlation in four radial bins. Including a correction for our selection function, we find the average environments of $z\simeq7.3$ quasars to be overdense within $4.5\,\rm{h}^{-1}\,\rm{cMpc}$. Assuming a power-law shape for the cross-correlation function with a slope of $\gamma=2.0$, we infer a median quasar-galaxy cross-correlation length of \rcrosscorr. 
    Under the assumption that quasars and galaxies trace the same underlying dark matter distributions and when adopting a galaxy auto-correlation length of $r_0^{\rm{GG}}=5$, we calculate a quasar auto-correlation length of \rautocorr.
    This auto-correlation length is lower than recent measurements at $z=6-7$ \citep[][]{Arita2023, Eilers2024, HuangJiamu2026, WangFeige2026} and provides tentative evidence for a nonmonotonic evolution of clustering properties at the highest redshifts (see top panel in Figure\,\ref{fig:clustering}).
    \item Utilizing predictions of a halo model framework, we use our quasar auto-correlation estimate to infer the minimum dark matter halo mass for $z\simeq7.3$ quasars to be \logMmin.
    Our estimate presents a departure from recent studies at $z=6-7$, which consistently find $M_{\rm{halo,min}}\approx10^{12}\,\rm{M}_\odot$ for UV-luminous quasars (see central panel in Figure\,\ref{fig:clustering}). However, we note that our estimate has significant statistical uncertainties and requires assumptions on the shape of the cross-correlation function and the galaxy auto-correlation length.
    \item Dark matter halos above our minimum halo mass have an abundance of \lognhalo, significantly above the expected number densities of high-redshift quasars.
    As a result, we calculate a quasar duty cycle of \fD, equivalent to a quasar lifetime of \tQ. 
    This value is consistent with independent constraints from quasar damping wings (bottom panel of Figure\,\ref{fig:clustering}) and suggests that growing SMBHs only appear as UV-luminous quasars for a small fraction of the Hubble time at $z\simeq7.3$.
\end{itemize}

Albeit limited by small number statistics, this work presents a first important step to investigate the environments and constrain the dark matter halo masses of quasars at the current redshift frontier.
Methodologically, the two-step (NIRCam photometry + NIRSpec/MSA spectroscopy) design of our observations presented challenges in galaxy selection and completeness analysis that are naturally circumvented with JWST/NIRCam wide-field slitless spectroscopy as demonstrated in \citet{Eilers2024}, \citet{HuangJiamu2026}, and \citet{WangFeige2026};
a valuable lesson learned for future studies of quasar environments with JWST.

The nonmonotonic behavior in clustering properties with redshift revealed by our analysis came as a surprise. 
To complete the full evolutionary picture (see Figure\,\ref{fig:clustering}) obtaining quasar-galaxy clustering measurements $z=4$--$6$ would be the next logical step. Importantly, those measurements should use the same large-scale-structure tracers, \oiiibr-emitting galaxies, at the same physical scales for a faithful comparison.

With the advent of the \textit{Euclid} Wide Survey \citep{EC:Scaramella022}, we should expect the discovery of many more $z>7$ quasars \citep{EC:Barnett2019} in the coming years.
This sample will provide the necessary statistical power to revisit our quasar-galaxy clustering analysis at $z\gtrsim7$ for much tighter constraints on the dark matter environments of the highest redshift quasars.

\begin{acknowledgements}
We thank the anonymous referee for their review of our manuscript. 
Furthermore, J.-T.S. would like to thank Jiamu Huang and Feige Wang for sharing the clustering results from the ASPIRE survey ahead of their publication.
J.-T.S. acknowledges funding by the Deutsche Forschungsgemeinschaft (DFG, German Research Foundation) - Project number 518006966 - and support by the DFG under Germany’s Excellence Strategy – EXC 2121 „Quantum Universe“ – 390833306.
This work has been further supported  by the DFG - Project number 506672582 (S.E.I.B.), NSF grants AST-2308258 (X.F.) and AST-2513040 (F.W.), and the Cosmic Dawn Center (DAWN), which is funded by the Danish National Research Foundation under grant DNRF140 (K.K.; DAWN Fellowship).
This paper includes data from the LBT. The LBT is an international collaboration among institutions in the United States, Italy, and Germany. The LBT Corporation partners are: The University of Arizona on behalf of the Arizona university system; Istituto Nazionale di Astrofisica, Italy; LBT Beteiligungsgesellschaft, Germany, representing the Max Planck Society, the Astrophysical Institute Potsdam, and Heidelberg University; The Ohio State University; The Research Corporation, on behalf of The University of Notre Dame, University of Minnesota and University of Virginia.
Some of the data presented herein were obtained at Keck Observatory, which is a private 501(c)3 non-profit organization operated as a scientific partnership among the California Institute of Technology, the University of California, and the National Aeronautics and Space Administration. The Observatory was made possible by the generous financial support of the W. M. Keck Foundation. 
The authors wish to recognize and acknowledge the very significant cultural role and reverence that the summit of Maunakea has always had within the Native Hawaiian community. We are most fortunate to have the opportunity to conduct observations from this mountain.
This work is based [in part] on observations made with the NASA/ESA/CSA James Webb Space Telescope. The data were obtained from the Mikulski Archive for Space Telescopes at the Space Telescope Science Institute, which is operated by the Association of Universities for Research in Astronomy, Inc., under NASA contract NAS 5-03127 for JWST. These observations are associated with program GO \#2073.
Support for program GO \#2073 was provided by NASA through a grant from the Space Telescope Science Institute, which is operated by the Association of Universities for Research in Astronomy, Inc., under NASA contract NAS 5-03127.
\end{acknowledgements}

\bibliographystyle{aa} 
\bibliography{all_new}

\begin{appendix}
\onecolumn
\section{Galaxy discoveries in the two quasar fields}
\begin{figure*}[h!]
    \centering
  \includegraphics[width=0.8\linewidth]{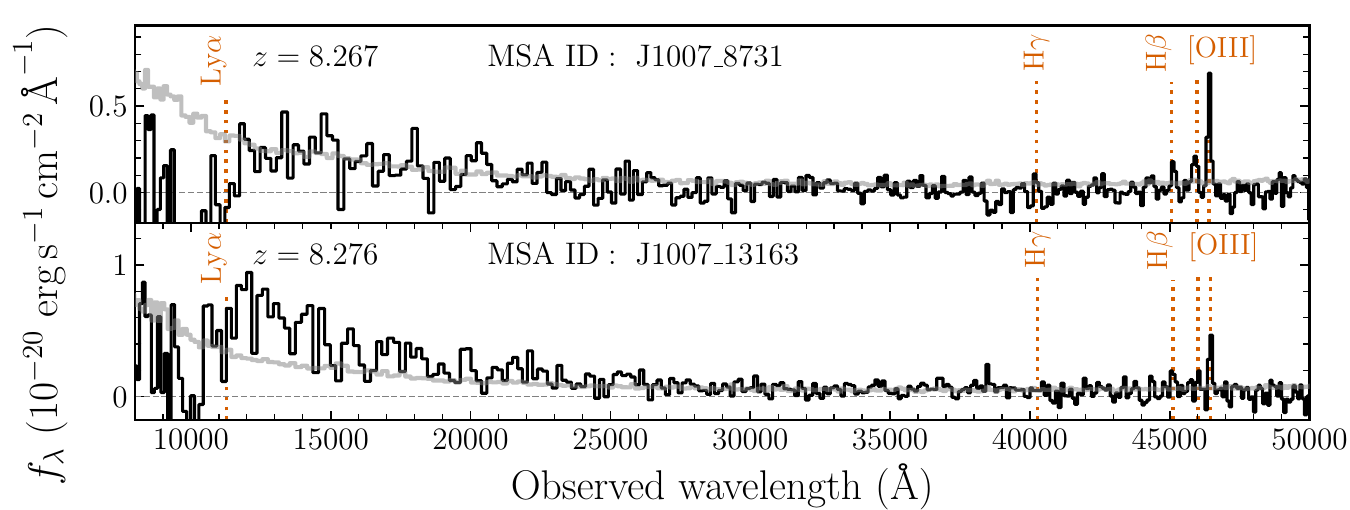}
    \caption{Discovery spectra of galaxies in the background of quasar J1007$+$2115 ($z=7.5149$). The spectrum is shown in black; the  vertical orange annotations highlight  possible emission line features and the position of the \Lya break. The uncertainties ($1\sigma$) on the spectral flux are shown in gray.}
    \label{fig:J1007_bg_gal_1D}
\end{figure*}

\begin{figure*}[ht!]
    \centering
 \includegraphics[width=0.8\linewidth]{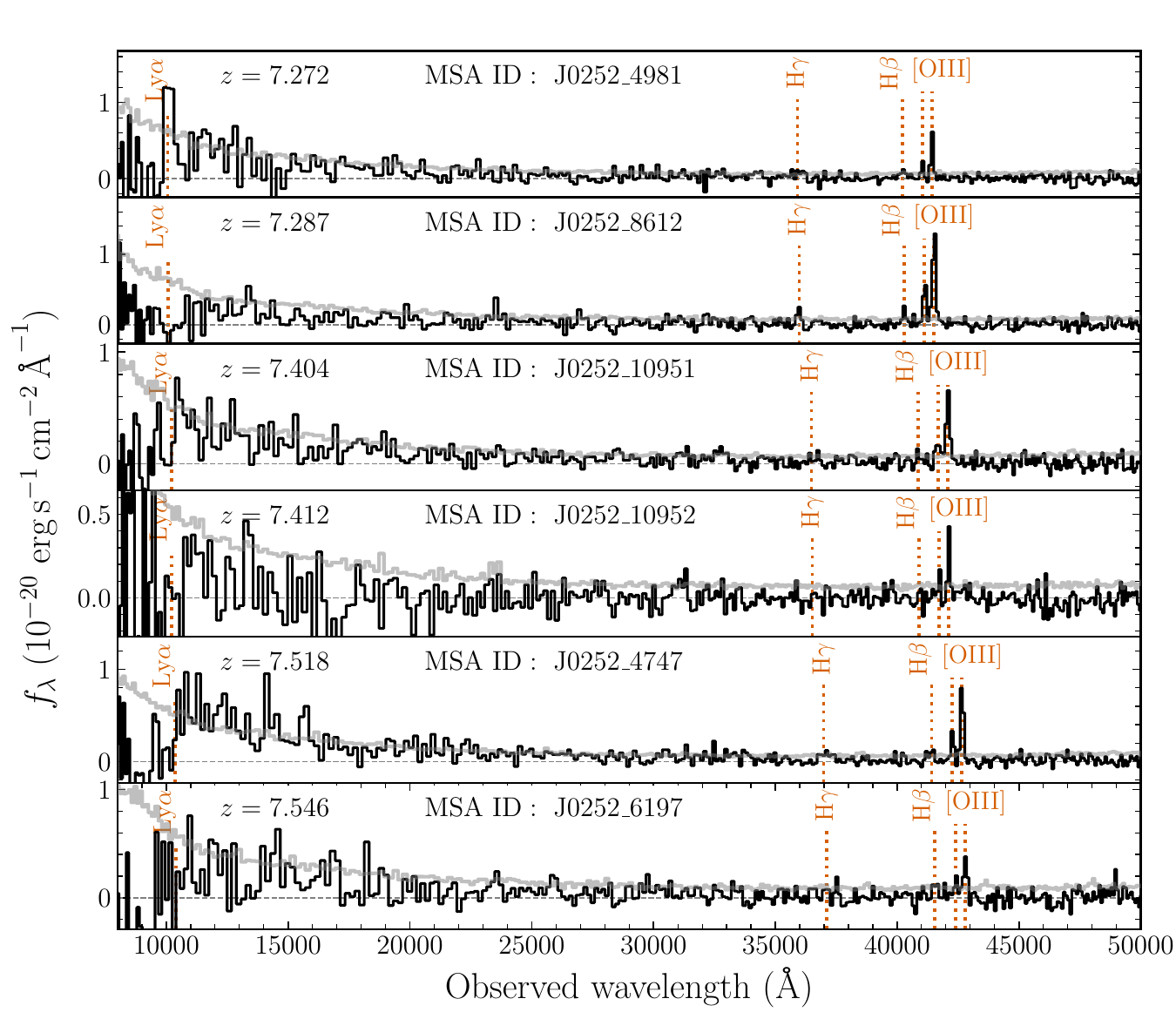}
    \caption{Discovery spectra of galaxies in the background of the quasar J0252$-$0503 ($z=7.00$). The spectrum is shown in black; the vertical orange annotations highlight  possible emission line features and the position of the \Lya break. Uncertainties ($1\sigma$) on the spectral flux are shown in gray.}
    \label{fig:J0252_bg_gal_1D}
\end{figure*}

In this appendix we provide additional figures displaying the discovery spectra of galaxies in the background of quasar J1007$+$2115 (Figure\,\ref{fig:J1007_bg_gal_1D}) and quasar  J0252$-$0503 (Figure\,\ref{fig:J0252_bg_gal_1D}).

\end{appendix}

\end{document}